\documentclass{jpp}
\usepackage{graphicx}
\usepackage{epstopdf, epsfig}
\usepackage{hyperref}
\usepackage{amsmath,amsfonts,amssymb}
\usepackage{subfigure}
\usepackage{natbib}
\usepackage{url}
\usepackage{bm}

\def\bfv{{\bf v}}

\def\bfB{{\bf B}}

\makeatletter
\def\fvec#1{\underline{\sbox\tw@{$#1$}\dp\tw@\z@\box\tw@}}
\makeatother

\renewcommand{\vec}[1]{\ensuremath{\bm{\mathit{#1}}}}
\newcommand{\ud}{\mathrm{d}} 

\shorttitle{Scale Statistics in Relativistic Turbulence}
\shortauthor{R.~F.~Serrano, J.~N\"attil\"a and V.~Zhdankin}

\title{Scale Statistics of Current Sheets in Relativistic Collisionless Plasma Turbulence}

\author{Roberto F. Serrano\aff{1}\corresp{\email{roberto.serrano32@lagcc.cuny.edu}}, Joonas\,N\"attil\"a\aff{2,3,4} and Vladimir Zhdankin\aff{5,4}}

\affiliation{\aff{1}Department of Natural Sciences, LaGuardia Community College, City University of New York, 31-10 Thomson Ave., Long Island City, NY 11101, USA
\aff{2}Department of Physics, University of Helsinki, P.O. Box 64, FI-00014, University of Helsinki, Finland
\aff{3}Physics Department and Columbia Astrophysics Laboratory, Columbia University, 538 West 120th Street, New York, NY 10027, USA
\aff{4}Center for Computational Astrophysics, Flatiron Institute, 162 Fifth Avenue, New York, NY 10010, USA
\aff{5}Department of Physics, University of Wisconsin-Madison, Madison, WI 53706, USA}

\begin{document}

\maketitle

\begin{abstract}
We analyze distributions of the spatial scales of coherent intermittent structures---current sheets---obtained from fully kinetic, two-dimensional simulations of relativistic plasma turbulence using unsupervised machine-learning data dissection. We find that the distribution functions of sheet length $\ell$ (longest scale of the analyzed structure in the direction perpendicular to the dominant guide field) and curvature $r_c$ (radius of a circle fitted to the structures) can be well-approximated by power-law distributions, indicating self-similarity of the structures. 
The distribution for the sheet width $w$ (shortest scale of the structure) peaks at the kinetic scales and decays exponentially at larger values. 
The data shows little or no correlation between $w$ and $\ell$, as expected from theoretical considerations. The typical $r_c$ depends linearly on $\ell$, which indicates that the sheets all have a similar curvature relative to their sizes. 
We find a weak correlation between $r_c$ and $w$. Our results can be used to inform realistic magnetohydrodynamic sub-grid models for plasma turbulence in high-energy astrophysics. 
\end{abstract}

\section{Introduction}

Over the last few decades, an increase in computational power has led to larger and larger direct numerical simulations of magnetohydrodynamic (MHD) turbulence 
\citep{2008matu.book.....B, Beresnyak2019, schekochihin_2022}. 
While these simulations generally support the theory of \citet{GS95} for self-similar fluctuations in MHD turbulence, the same simulations also show the presence of transient coherent structures---spatio-temporal fluctuations known as the intermittency \citep[see, e.g.,][for a review]{vlahos_etal_2023}. 
Among these coherent structures are current sheets, which are ribbon-shaped regions of high electric current density. 
Current sheets have also been identified in simulations of collisionless plasmas where the full kinetic equations are solved \citep[e.g.,][]{comisso&sironi2018, comisso&sironi2019}. 

Several groups have performed a statistical study of current sheets in MHD turbulence to categorize their geometric distributions \citep{servidio_etal_2010, uritsky_etal_2010, 2013ApJ...771..124Z, zhdankin_etal_2014, zhdankin_etal_2016, 2017MNRAS.467.3620Z, ross_latter_2018}. They measured the distribution of the width, length, and thickness of the current sheets. 
The results revealed that the current sheets are indeed self-similar structures, and their geometric components obey power-law or exponential distributions. In addition, recent works have detected and analyzed current sheets in collisionless regimes \citep{makwana_etal_2015, azizabadi_etal_2021, sisti_etal_2021, 2021SPIC...9916450B}. Others have also attempted to associate the properties of current sheets with the statistical distribution of turbulent field fluctuations, including recent works on relativistic MHD \citep{Chernoglazov_2021} and kinetic turbulence \citep{Davies2024}.

Here, we focus on current sheets in fully kinetic simulations of magnetically dominated (relativistic) plasma turbulence.
In the magnetically dominated regime, a large reservoir of magnetic energy is available to cascade down and dissipate, leading to heating and acceleration of non-thermal particles. 
None of the previous studies have performed the geometric statistical analysis in this regime of turbulence, which has $v_{A} \approx c$, where $v_{A}$ is the Alfv\'en velocity and $c$ is the speed of light. 
Relativistic kinetic turbulence is an area of active study in recent years, spurred by progress with particle-in-cell (PIC) simulations \citep{Zhdankin2017_prl,  comisso&sironi2018, 2021ApJ...921...87N, vega_etal_2022, meringolo_etal_2023, singh_etal_2024}. 
It is relevant for many high-energy astrophysical systems (e.g., compact objects and transients). 
The motivation of this study is to perform a 2D statistical analysis of current sheets and compare/contrast with previous results in the literature.  

In this article, we apply the computer vision algorithm developed by \cite{2021SPIC...9916450B} to 2D fully kinetic (PIC) simulations of relativistic turbulence in pair plasmas. 
Notably, we measure the distributions of spatial scales of current sheets in the relativistic regime of kinetic turbulence. 
Section~\ref{sect: theory} describes the fundamental timescales and length scales and presents analytical expectations for the statistical scalings. In Section~\ref{sect: methods}, we discuss the numerical setup of our PIC simulations, review the machine learning segmentation algorithm, and define our spatial scale measurements. In Section~\ref{sect:results}, we present our measured distributions and their scaling indices. 
In Section~\ref{sect:Discussion}, we discuss the similarities between our scaling indices and previous studies and the implications for high-energy astrophysical plasmas.

\section{Plasma length- and timescales}\label{sect: theory}

We analyze the properties of current sheets in collisionless electron-positron pair plasmas.
We consider pair plasma with an average temperature $\langle T \rangle$ (angular brackets $\langle \cdot \rangle$ indicate spatial averages in the remainder of the paper). 
Then, the average dimensionless temperature is given by
\begin{equation}
 \theta  \equiv \frac{k_{\mathrm{B}} 
     \langle T \rangle}{m_\mathrm{e}c^{2}} \, ,
\end{equation}
where $k_{\mathrm{B}}$ is the Boltzmann constant, and $m_\mathrm{e}$ is the electron mass, and $c$ is the speed of light. 

The plasma is magnetized by threading it with a uniform background field $\mathbf{B}_0 = B_{0} \mathbf{\hat{z}}$. 
The dimensionless parameter quantifying the strength of the magnetic field is the magnetization, given by
\begin{equation}
    \sigma_{\mathrm{0}} \equiv \frac{B_{\mathrm{0}}^{2}}{4\pi n_0m_\mathrm{e}c^{2}} \, ,
\end{equation}
where $n_{\mathrm{0}}$ is the total particle number density.

The microscopic timescale of the plasma is characterized by the (cold) plasma frequency,
\begin{equation}
    \omega_{\mathrm{p}} \equiv \bigg(\frac{4\pi n_0 e^2 }{m_\mathrm{e}}\bigg)^{1/2} \, ,
\end{equation}
where $e$ is the electron charge. In relativistic plasmas, the plasma frequency also defines the microscopic length scale of the system, known as the plasma skin depth,
\begin{equation}
    \frac{c}{\omega_{\mathrm{p}}} = \bigg(\frac{m_\mathrm{e} c^2}{4\pi n_0 e^2}\bigg)^{1/2} \, .
\end{equation}

The  (non-relativistic) gyro-frequency is
\begin{equation}
    \label{eqn: gyrofrequency}
    \omega_B = \frac{e B_0}{m_\mathrm{e} c} \, .
\end{equation}

The related gyroradius is
\begin{equation}
    \label{gyroradius}
    r_g \equiv \frac{c}{\omega_B} \frac{\langle p \rangle}{m_\mathrm{e} c} =\sigma_0^{-1/2}  \frac{c}{\omega_{\mathrm{p}} } \frac{\langle p \rangle}{m_\mathrm{e} c} \, ,
\end{equation}
where $p$ is the momentum of the particle.
In the following, we use the skin depth $c/\omega_{\mathrm{p}}$ as the reference scale when measuring sizes. This choice is made because $c/\omega_{\mathrm{p}}$ is the largest of the kinetic scales when $\sigma_0 > 1$, thus determining the characteristic width of structures.

\subsection{Current Sheet Distributions}

The predominant intermittent structures in magnetically-dominated plasma turbulence are current sheets.%
\footnote{Apart from the electric current, $\mathbf{J} \propto \nabla \times \bfB$, vorticity of the velocity field, $\vec{\zeta} \equiv \nabla \times \bfv$, can be used to characterize the local ``rotation'' of the flow. 
In analog to current sheets, intermittent vorticity sheets exist in plasma turbulence. We do not discuss vorticity sheets here since we focus on magnetically-dominated plasmas.}
In MHD turbulence, these structures are responsible for a large fraction of the overall resistive dissipation \citep[e.g.,][]{zhdankin_etal_2016}. 
In collisionless plasmas, they are likewise thought to be sites of energy dissipation through either magnetic reconnection or other kinetic damping mechanisms \citep[e.g.,][]{wan_etal_2012, tenbarge_etal_2013, howes_2016, parashar_matthaeus_2016, loureiro_boldyrev_2017, mallet_etal_2017, tenbarge2021b, nattila2022b, borgogno_etal_2022, Chernoglazov2023, Comisso2023, Davies2024}. 

\begin{figure}
    \centering
    \includegraphics[trim={10.0cm 0.0cm 10.0cm 0cm}, clip=true, width = 0.45\columnwidth]{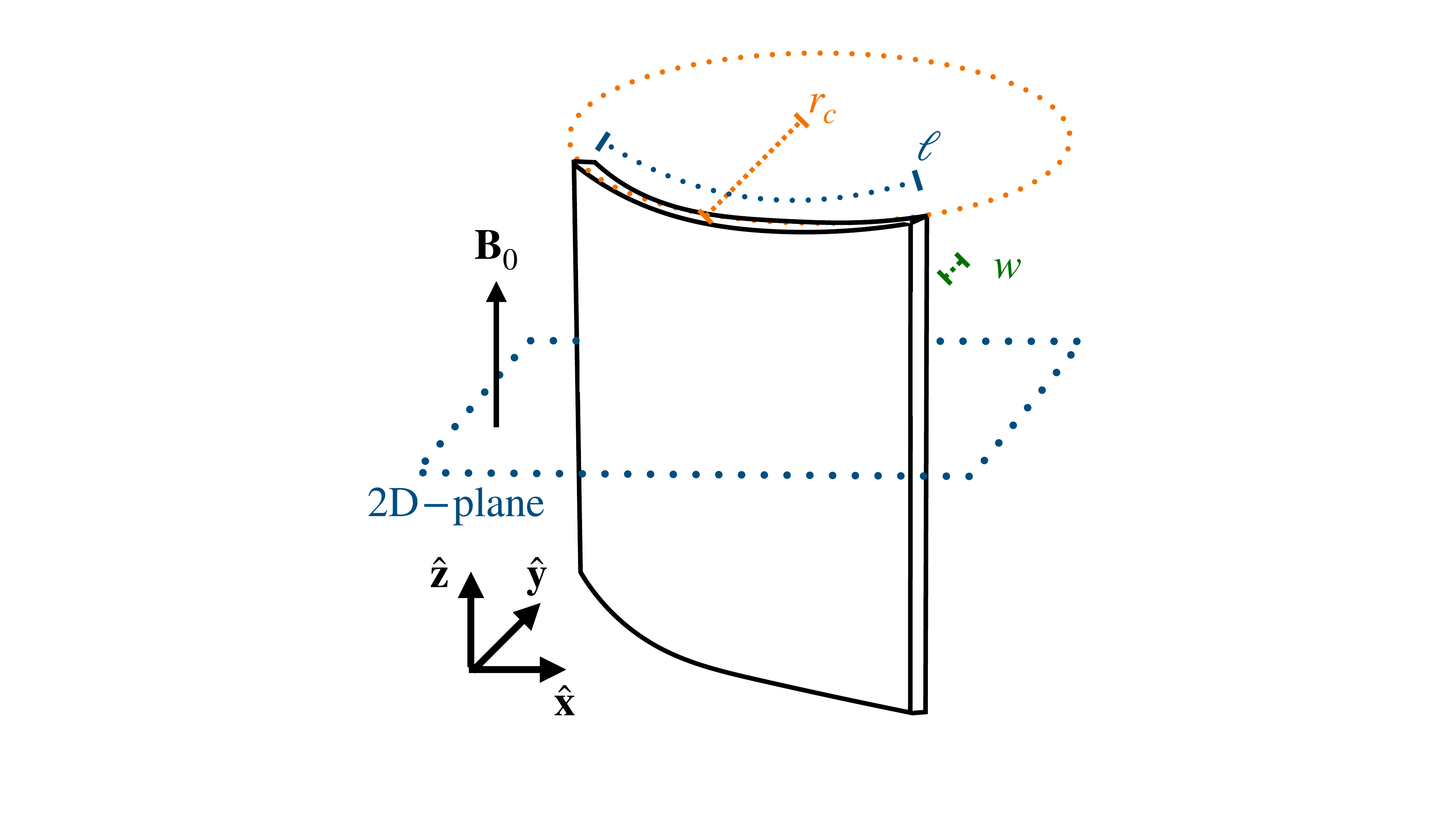}
    \caption{Schematic illustration of a three-dimensional current sheet along a background magnetic field $\mathbf{B}_0$. 
    The two directions perpendicular to the magnetic field are denoted with the 2D plane.
    The length $\ell$ of a sheet (blue curved segment), followed by the width of the sheet $w$ (green line), and the radius of curvature $r_c$ (orange line).}
    \label{fig:fig sheet}
\end{figure}

In domains with the full three spatial dimensions (3D), current sheets have a ribbon-like morphology and, therefore, can be ideally characterized by three
scales (ignoring curvature and finer structure). 
For two-dimensional (2D) simulations with an out-of-plane background magnetic field $\mathbf{B}_0$, the largest scale is in the direction of $\mathbf{B}_0$, which we will ignore. 
The remaining two scales are in the directions perpendicular to the background magnetic field; we use the terminology \emph{length} $\ell$ and \emph{width} $w$ to describe these two scales, ordered such that $\ell > w$. See Fig.~\ref{fig:fig sheet} for a schematic view of our general definitions. We emphasize that this differs from the terminology used in 3D studies. 

Numerical simulations of MHD turbulence exhibit current sheets that have a broad distribution of possible values of $\ell$, spanning the turbulent inertial range; in particular, the data is well-described by a power-law distribution when $\ell$ is in the inertial range:
\begin{equation}
 P_\ell(\ell) \equiv \frac{\ud N_\mathrm{sheets}}{\ud \ell} \propto \ell^{-\alpha} \, ,
\end{equation}
where $N_\mathrm{sheets} = \int P_\ell(\ell) \, \ud \ell$ is the number of sheets, and $\alpha \approx 3.3$ is measured in MHD turbulence simulations \citep{zhdankin_etal_2016}. 
We note that $\alpha$ is used as a general index for any power law distribution in the following text.
The widths, on the other hand, mainly reside near the dissipation scale $s$, with a narrowly peaked distribution, 
which we will approximate as
\begin{equation}
P_w(w) \equiv \frac{\ud N_\mathrm{sheets}(w)}{\ud w} \propto \exp\left(- \frac{w}{s} \right) \, ,
\end{equation}
where $s \sim c/\omega_{\mathrm{p}}$.
Furthermore, we measure the circular curvature of the current sheets, $r_c$, which may be expected to be a power-law distribution, 
\begin{equation}
 P_{r_c}(r_c) \equiv \frac{\ud N_\mathrm{sheets}}{\ud r_c} \propto r_c^{-\alpha} \, ,
\end{equation}
because the structures are distorted by turbulent eddies spanning a broad range of scales. The power-law index for $P_{r_c}(r_c)$ is nontrivial to predict from theory. 
In addition, we will be analyzing the joint 2D distributions 
\begin{equation}
P_{w,\ell}(w,\ell) \equiv \frac{\partial^2 N_\mathrm{sheets}(w, \ell)}{\partial w \, \partial \ell} \, ,
\end{equation}
\begin{equation}
P_{w, r_c}(w, r_c) \equiv \frac{\partial^2 N_\mathrm{sheets}(w, r_c)}{\partial w \, \partial r_c} \, ,
\end{equation}
and
\begin{equation}
 P_{r_c, \ell}(r_c,\ell) \equiv \frac{\partial^2 N_\mathrm{sheets}(r_c, \ell)}{\partial r_c \, \partial \ell}   \, ,
\end{equation}
that capture the correlations between $\ell$, $w$, and $r_c$.
The 1D distributions follow naturally from the marginalization of these distributions as, for example, $P_w(w)= \int P_{w,\ell}(w,\ell) \, \ud \ell$. 
We anticipate that since $w \sim s \sim c/\omega_{\mathrm{p}}$, the width will be weakly correlated (if at all) with the other scales. 
The correlation between $r_c$ and $\ell$ is nontrivial to predict, in general, since structures of varying lengths will be distorted by different populations of turbulent eddies. 
However, we note that $r_c \propto \ell$ if the relative curvature of the structure is scale independent (i.e., the typical shape of a curved structure is independent of its length).

\section{Methods}\label{sect: methods}
\subsection{Numerical Methods and Setup}\label{sec: 2.1}

\begin{figure*}
\centering
    \includegraphics[width=0.45\columnwidth]{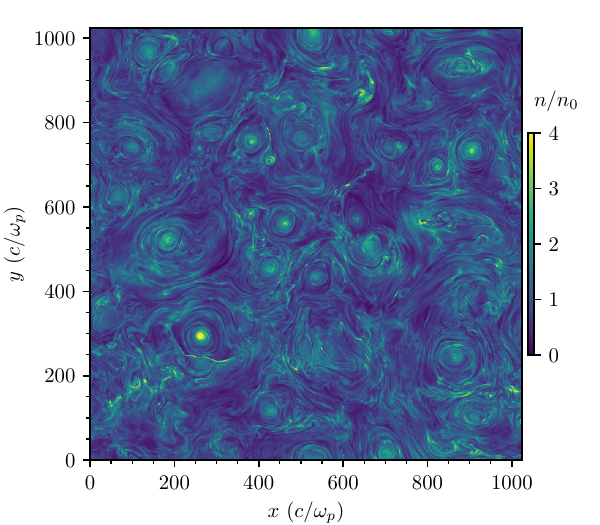}
    \includegraphics[width=0.45\columnwidth]{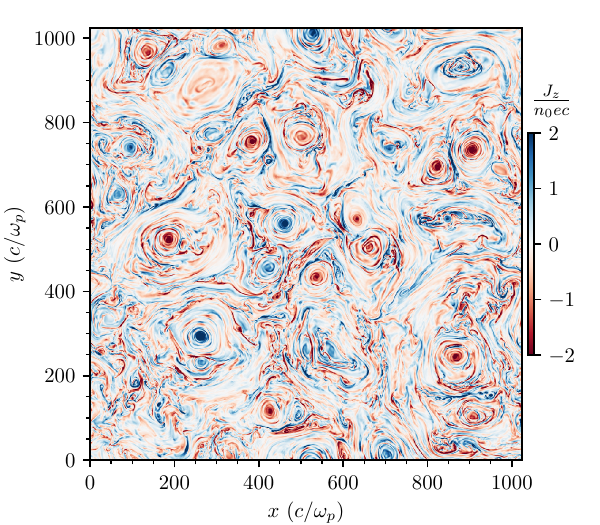}
    \includegraphics[width=0.45\columnwidth]{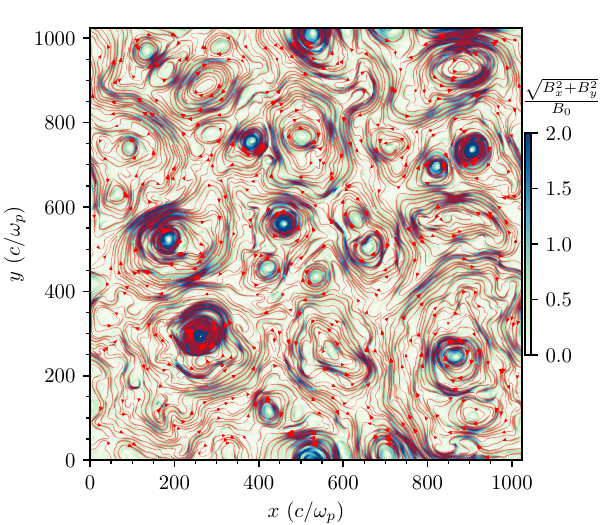} 
    \includegraphics[width=0.45\columnwidth]{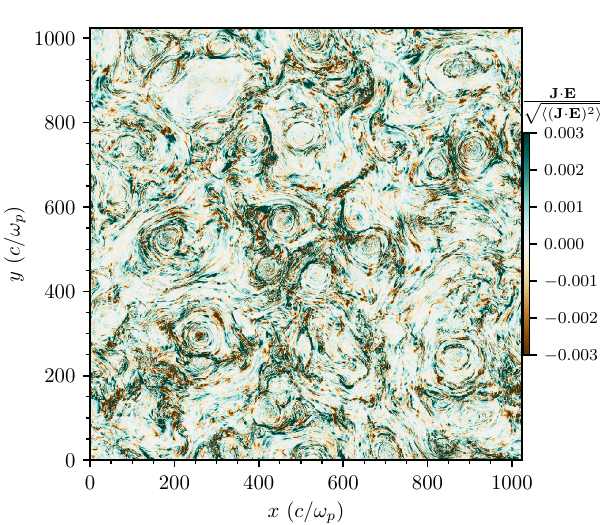}
    \includegraphics[width=0.45\columnwidth]{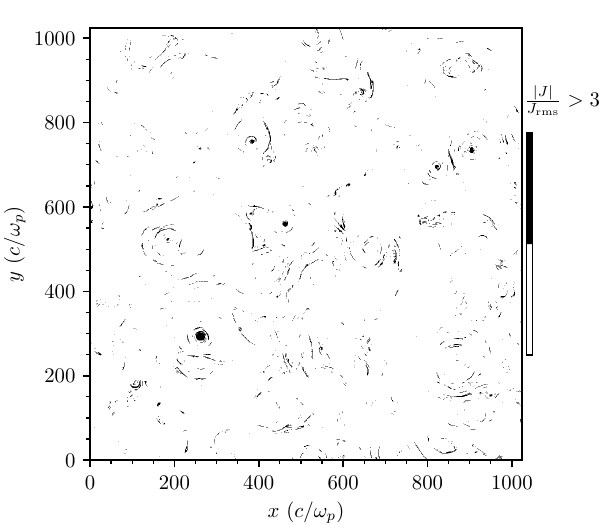}
    \includegraphics[width=0.45\columnwidth]{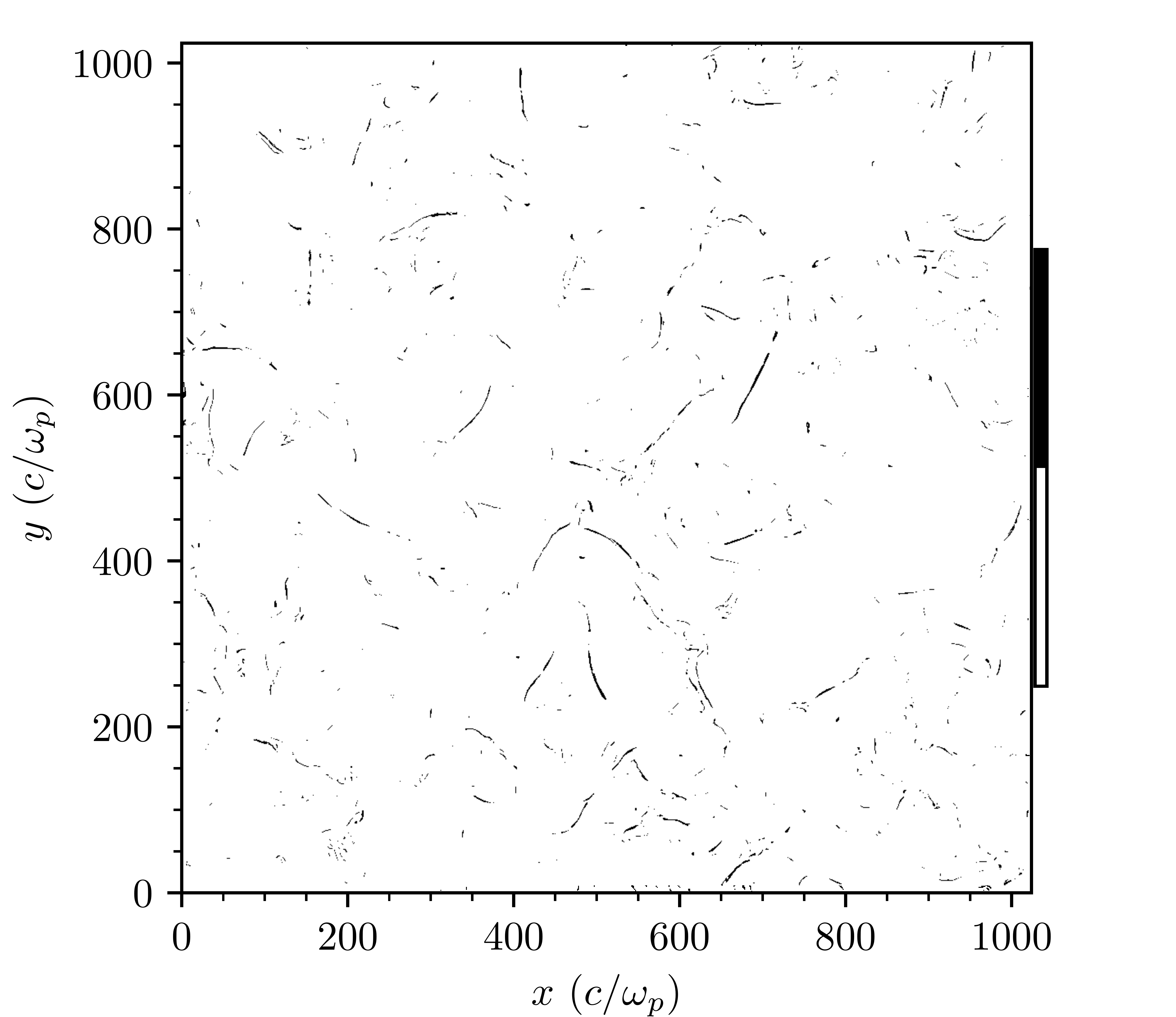}    
\caption{Visualization of the analyzed turbulence data measured at time $t = 4.6 l_0/c$: 
plasma density $n/n_0$, where $n_0$ is the initial plasma density (upper-left panel);  
out-of-the-plane current density $J_{z}/n_0 e c$ (upper-right panel);
strength of the in-plane magnetic field component $\sqrt{B_x^2+B_y^2}/B_0$, and the field lines (red curves), where $B_0$ is the initial guide field strength (middle-left panel);
a proxy of the work done by the parallel electric field $\mathbf{J} \cdot \mathbf{E}/\sqrt{\langle (\mathbf{J} \cdot \mathbf{E})^2 \rangle}$ in units of the rms-value (middle-right panel);
regions of the current density with $J/J_{\mathrm{rms}} > 3$, where $J_\mathrm{rms} = \sqrt{\langle J^2 \rangle}$ (bottom-left panel); and,
current sheet regions from the SCE algorithm (bottom-right panel). 
The SCE algorithm is shown at $t = 4.0 l_0/c$.
} 
\label{fig:1}
\end{figure*}

We analyze the two-dimensional freely evolving (decaying) turbulence simulations presented in \citet{2021ApJ...921...87N}. Here, we briefly review the simulation setup, and initial conditions, and refer the reader to the original paper for more details. 

The simulations are initialized with a homogenous pair plasma plasma with a uniform density $n_{\pm,0}$ and an initial temperature $\theta_{\mathrm{0}} = 0.3$.
The plasma is threaded by a uniform out-of-plane magnetic field $\mathbf{B}_0 = B_0 \mathbf{\hat{z}}$. To seed the turbulence, we perturb the magnetic field in the directions perpendicular to the magnetic field, $\mathbf{B}_{\perp} = (B_{\mathrm{x}}, B_{\mathrm{y}},0)$ with sinusoidal large-scale modes with wavelength $l_0 = L/8$ (where $L$ is the size of the full simulation domain).

The simulations use the relativistic PIC module in \textsc{runko} framework \citep{nattila2022c}. 
The PIC algorithm evolves all three components of the electromagnetic fields using a second-order finite-difference scheme, while particles evolve in time using a relativistic Boris pusher. 
In addition, we impose doubly periodic boundary conditions on the domain and perform 8 passes of current filtering between each step. 

The 2D domain in our simulations is a square in the x-y plane of size $L = 1024 c/\omega_{\mathrm{p}}$, and it is covered with $5120^{2}$ cell resolution. 
The (initial) plasma skin depth is resolved by $5$ grid cells. 
In our 2D simulations, we initialize fluctuations on a scale of $l_{0} \approx 125 c/\omega_{\mathrm{p}}$. 
We simulate the turbulence for $20 l_{0}/c$. 
We focus on two simulations with $\sigma_0 = 1$ and $10$, which correspond to the relativistic (magnetically dominated) regime. 
During the simulation, the plasma heats up and so $\theta_{0}$ increases (see fig.~6 in \citealt{2021ApJ...921...87N}).
 For our simulation with $\sigma_0 = 1$ at $t = l_{0}/c$, we measure $\theta \approx 0.45$ which then increases to $\theta \approx 0.60$ at $t = 15l_{0}/c$. 
Similarly, for the case of $\sigma_0 = 10$, we first measure $\theta \approx 0.50$ which then increases to $\theta \approx 2.70$. 
Thus, the temperatures become only mildly relativistic, and it is reasonable to use the non-relativistic expressions for kinetic scales from Sec.~\ref{sect: theory}. 

We present six visualizations of one of our simulations with $\sigma_0 = 10$ at $t = 4.6\, l_0/c$ in Fig. \ref{fig:1}.
The magnetic eddies are visible for the fully developed turbulence as large plasma over-densities (upper-right panel). 
Secondly, we can visually trace out two types of structure in the out-of-the-plane current density component $J_z$ (upper-left panel): elongated stripes (current sheets) and circular spots (plasmoids). 
These same structures are also visible as coherent structures in the (perpendicular) magnetic field $\mathbf{B}_\perp$ (middle-left panel): the current sheets coincide with regions of alternating magnetic fields (in between colliding plasmoids). The plasmoids coincide with the location of magnetic loops. The regions with current sheets (with anti-parallel fields) are prominent locations of energy dissipation. 
The sheet regions can be associated with areas of energy dissipation when visualizing the work done by the electric field $\propto \mathbf{J}\cdot\mathbf{E}$ (middle-right panel). 
We note that a significant fraction of the regions with high electric currents have magnetic fields that are parallel (rather than anti-parallel) on both sides (Ha, N\"attil\"a, Davelaar, Sironi, 2024, \textit{in prep}).

A standard method to segment the current sheets into individual structures is to apply a threshold in the current density.
To illustrate this method, we trace out the outlines of the sheets by applying a threshold of $J_{\mathrm{z}}/J_{\mathrm{rms}} > 3$ (bottom-left panel).
In our case, however, we use the regions originating from a more advanced Region-Of-Interest (ROI) detection (measured at a time $4l_0/c$), relying on a computer vision segmentation algorithm presented in \citet{2021SPIC...9916450B} (bottom-right panel).
Such segmentation method has the advantage that it clusters the sheets based on not only the electric current but also other features;
this enables us to separate the current sheets and plasmoid cores from each other.
The details of the method are given in the next section.

\begin{figure*}
    \centering
    \includegraphics[width=0.95\textwidth]{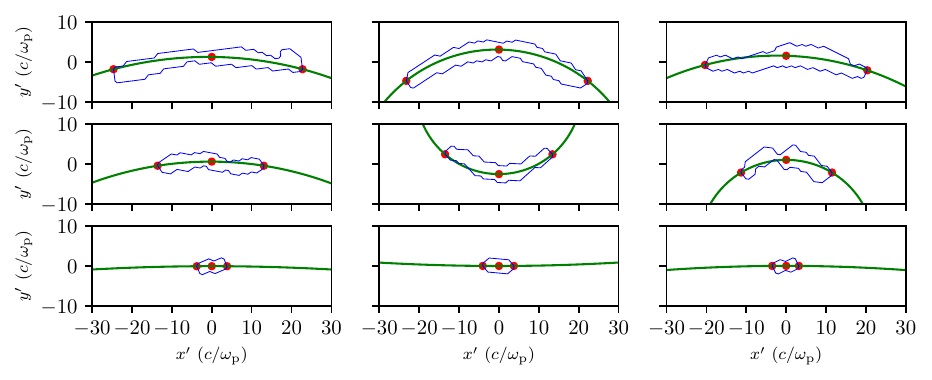}
    \caption{
    Examples of current sheets in our catalog.
    Individual current sheets of varying length, $\ell$, and radius of curvature, $r_{c}$, are visualized in the transformed position space $(x',y')$ using Eqs.~\ref{eqn: rotation matrix},~\ref{eq: primecoordinates}.
    The three red points mark the locations used to calculate the curvature radius using Eq.~\ref{eqn: complex z1}. 
    The green line shows the arc of the circle passing through the three points, with the center calculated using Eq.~\ref{eqn: complex center org}.}
    \label{fig: sheet gallery}
\end{figure*}

\subsection{Overview of the Machine Learning Algorithm}
We employ the unsupervised ensemble machine learning algorithm developed by \citet{2021SPIC...9916450B} to segment the current sheets from fully kinetic plasma turbulence simulations. 
The algorithm is an ensemble extension of the Self Organizing Map (SOM) algorithm \citep{Kohonen2001, Kohonen2013} dubbed ``Statistically Combined Ensemble" (SCE), which merges multiple (independent) SOM evaluations together, hence increasing the robustness of the ROI boundaries.

Briefly, we outline how the algorithm works. 
We apply the SOM algorithm to our simulation snapshots and cluster the data using three features: 
$\bold{B_\perp}$, $\vec{J_z}$, and $\vec{J_z} \cdot \vec{E}$, where $\bold{B_\perp}$ is the (in-plane) perpendicular magnetic field, $\vec{J_z}$, is the (out-of-the-plane) parallel component of the current, and $\vec{J_z} \cdot \vec{E}$ is a measure of the work done by the electric field $\vec{E}$. 
Each pixel on the image then defines a multidimensional feature-space data point.

The SOM algorithm is based on a 2D network of neurons that are trained to approximate the multidimensional (feature-space) data.
In practice, the map can be thought of as a (finite) surface that is fitted to the input data points.
Conversely, each feature-space data point can be associated with a best-matching neuron on the map.
The map defines a latent (reduced) data space for the input data.

The neural map is then used to cluster the feature-space data into different groups by associating a region on the map (i.e., a set of connected neurons) as one data cluster.
The (feature-space) distance between the neurons defines a convenient cost function to separate the neurons into a finite number of clusters.
In theory, the number of resulting clusters is arbitrary. 
Still, in practice, the map often has distinct regions of closely connected neurons that help set the number of resulting clusters from the analysis.
In our case, the SOM algorithm results, on average, in 4 distinct clusters, one of which is the current sheets \citep[see,][for more details]{2021SPIC...9916450B}.%
\footnote{The clustering analysis also identifies plasmoids as a separate cluster. Here, we omit the analysis of that cluster because the chosen 2D geometry makes the plasmoids longer-lived compared to 3D.}.

In practice, we use an ensemble version of the SOM to segment the sheets.
This so-called SCE method combines multiple SOM segmentations into one clustering realization to increase the signal-to-noise ratio of each ROI boundary.
In general, our ML-based segmentation method should be contrasted with the previously used threshold technique, where the boundaries of the current sheets are found by setting a threshold value for the magnitude of the electric current (see the lower right panel in Fig. \ref{fig:1}).
We emphasize that our definition of a current sheet is more complex than what results, e.g., from segmenting the sheets with the threshold method:
our definition of the current sheets is based on a combination of all three input features.
The second reason, in addition to the more generic definition of the current sheet cluster,

is that the SCE method is computationally faster, making it feasible to construct large structure catalogs. 
We note, however, that the algorithm is currently calibrated only on 2D data, so we focus on that here.

\subsection{Measurements}

The following section presents our definitions of geometric length and width.
Specifically, we analyze a time series of 2D masks separated by $\Delta t = 1 l_0/c$ for two simulations with $\sigma_0 = 1$ and $\sigma_0 = 10$. 
We analyze a time interval, $l_0/c \leq t \leq 15 l_0/c$ when the turbulence is fully developed but still sustains large amplitude fluctuations.
In practice, when analyzing the structures, we use the open-source geometry library  \texttt{shapely} \citep{shapely} to automate the geometric analysis of the structures.
First, we dissect each independent closed contour in a mask into $(x,y)$ coordinate lists that trace the element's boundary. This procedure is repeated for every mask in the time series.
Specifically, we use the \texttt{shapely.Polygon} class, which takes a set of $N_p$ data points and connects the chain of coordinate points into a ``ring''.
We generate an instance of this polygon class for each structure in a given image. 
Once we generate the \texttt{Polygon} class to represent the current sheet, we measure its perimeter $\mathcal{S}$ ($=\oint \ud s$, where the line integral is performed along the closed curve around the polygon) and use it to define a robust estimator of the sheet length $\ell$ as
\begin{equation}\label{eqn: perimeter}
    \mathcal{S} \approx 2\ell + 2 w \approx 2\ell,
\end{equation}
so that $\ell \approx \mathcal{S}/2$, where we assumed that $w \ll \ell$ in the second part of the equation. 
We also tried measuring $\ell$ by taking the pair-wise distance between each coordinate point and identifying the largest distance as $\ell$. 
However, we forgo this definition since we find it to be a more noisy estimator of $\ell$. We also note that when accounting for the possible curvature of the structure, the natural length is the arc length of the structure, not the distance along the structure's longest dimension; therefore, our definition $\ell$ also naturally includes the curvature effects.

To measure the width (see Appendix ~\ref{app:1} for the full derivation), we take each structure and calculate the major axis of each sheet, which is defined as the largest distance between any set of points. Then, we rotate the structure such that its major axis lies on the $x$ axis. Then, we translate the structure to the origin. Once in this new reference frame, we measure the width using Eq.~\ref{eqn: width}. 

Finally, we introduce a new measurement of the curvature of the current sheet, dubbed ``curvature radius", $r_{c}$. 
The radius of curvature is calculated by fitting a circle through the 3 points on the curve: leftmost tip, center point, and rightmost point (in the new reference frame described above). Then, a circle with a radius $r_c$ can be defined to pass through each point. In the limit that the curvature radius $r_c \rightarrow 0$, the structure is highly curved, and as $r_c\rightarrow \infty$, the structure becomes flat. We detail our simple circle fitting algorithm in the Appendix \ref{app:3}.

In Fig.~\ref{fig: sheet gallery}, we display representative examples from the resulting current sheet catalog for the simulation with $\sigma_0 = 10$. 
These examples show that most of the sheets have approximately constant width $w$. 

Lastly, the approximation of the sheet curvature as part of a circle's arc seems appropriate in almost all cases;
in total, we find that only about $0.5\%$ of the sheets can not be adequately modeled as arcs (but are, e.g., $S$- or $L$-shaped). 
Such non-standard morphologies can sometimes result from the interactions and mergers of sheets.

\section{Results}\label{sect:results}

Our catalog contains $\approx\!13\,000$ individual objects for $\sigma_0 = 1$ and $\approx\!10\,000$ for $\sigma_0 = 10$ simulations.
Each time snapshot contains, on average, $\approx\!1\,000$ current sheets.

\begin{figure}
\centering
    \label{fig: threshold plot}
    \includegraphics[width = 0.45\columnwidth]{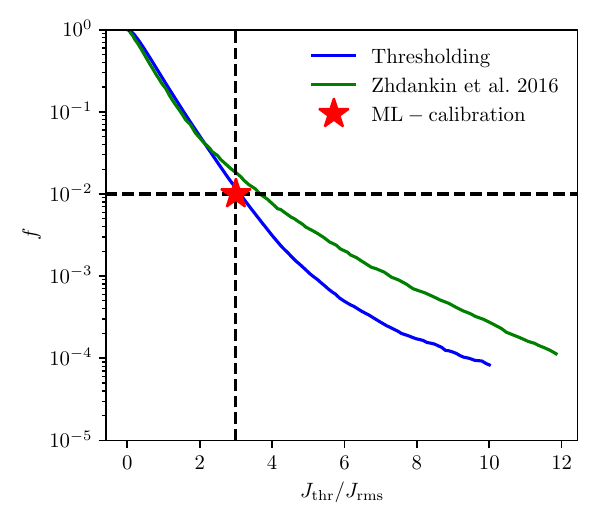}
    \caption{The filling fraction of current sheets, $f$, of our simulation of $\sigma_0 = 10$ measured at $4.6l_0/c$, as function of $J_{\mathrm{thr}}/J_{\mathrm{rms}}$ (blue line), where $J_{\mathrm{thr}}$ is the threshold factor, and where $J_\mathrm{rms} = \sqrt{\langle J^2 \rangle}$ is the root mean square of the current density. The star symbol denotes the filling fraction measured using our machine-learning algorithm. We also fit the MHD data from \citet{zhdankin_etal_2016} (green line).
    }
    \label{fig:threshold}
\end{figure}

We calibrate our current sheet identification method with the previous method of ``thresholding" used in \citet{zhdankin_etal_2014, zhdankin_etal_2016}. In Fig. ~\ref{fig: threshold plot}, we plot the volume filling fraction $f$ of the structures occupying pixels having current densities $|\mathbf{J}| > J_\mathrm{rms}$, where $J_\mathrm{rms}$ is a threshold parameter that we vary. 
The filling fraction of the structures with current densities above the threshold exhibits a steep exponential decay at low $J_{\rm thr}/J_{\mathrm{rms}} < 3$, which then flattens to a less steep decay for $J_{\mathrm{thr}}/J_{\mathrm{rms}} > 3$. 
We also compare our results to the MHD analysis by \citet{zhdankin_etal_2016} which shows a very similar behavior for $J_{\mathrm{thr}}/J_{\mathrm{rms}} > 3$. We find that the filling fraction of structures detected by the SCE algorithm agrees with the thresholding method when $J_{\mathrm{thr}}/J_{\mathrm{rms}} \approx 3$ (the red star in Fig. ~\ref{fig: threshold plot}). 
This indicates the ``effective" threshold of the SCE algorithm.

\begin{figure*}
\centering
    \includegraphics[width=0.45\columnwidth]{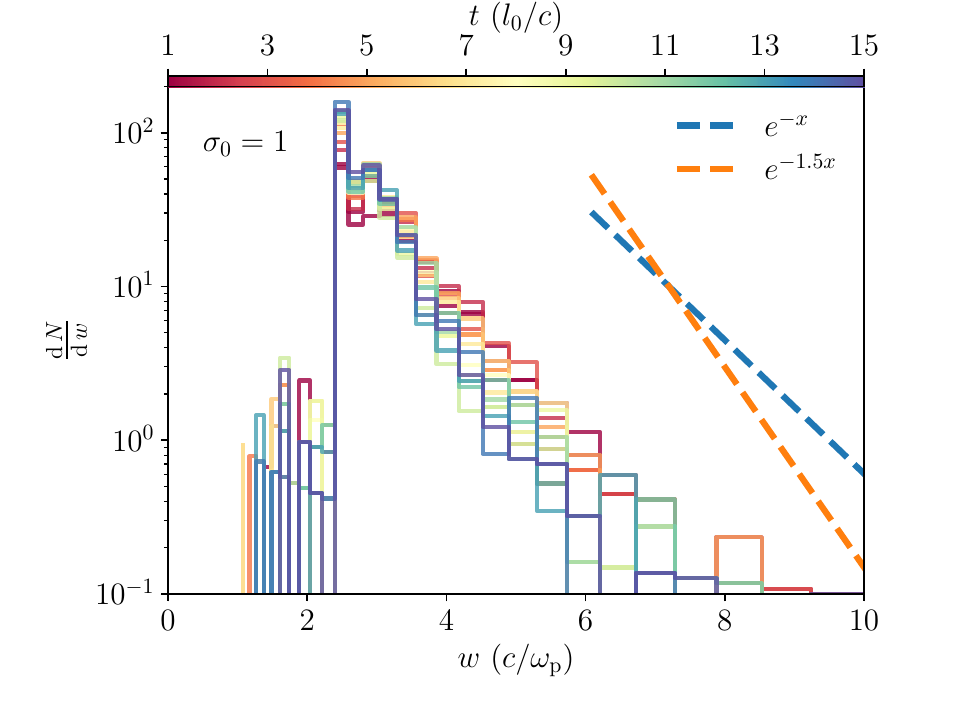}
    \includegraphics[width=0.45\columnwidth]{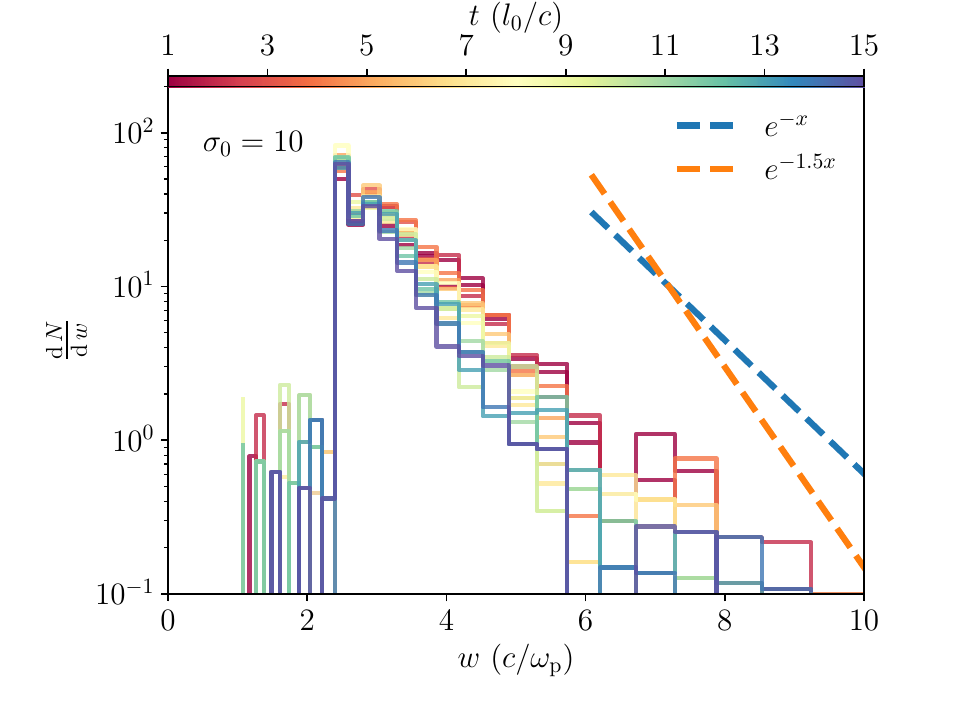}
    \caption{Width distributions, $dN/dw$, for simulations with initial $\sigma_0 = (1, 10)$. Two reference exponential fits with the corresponding index are shown in the legend.}
    \label{fig: width dist}
    \includegraphics[width=0.45\columnwidth]{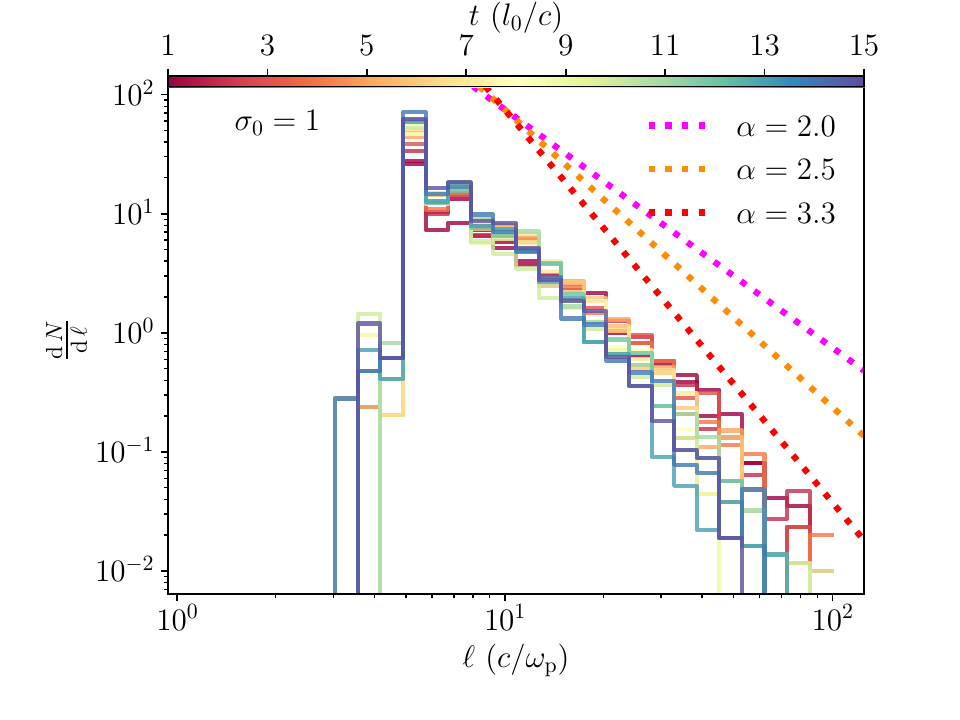}
    \includegraphics[width=0.45\columnwidth]{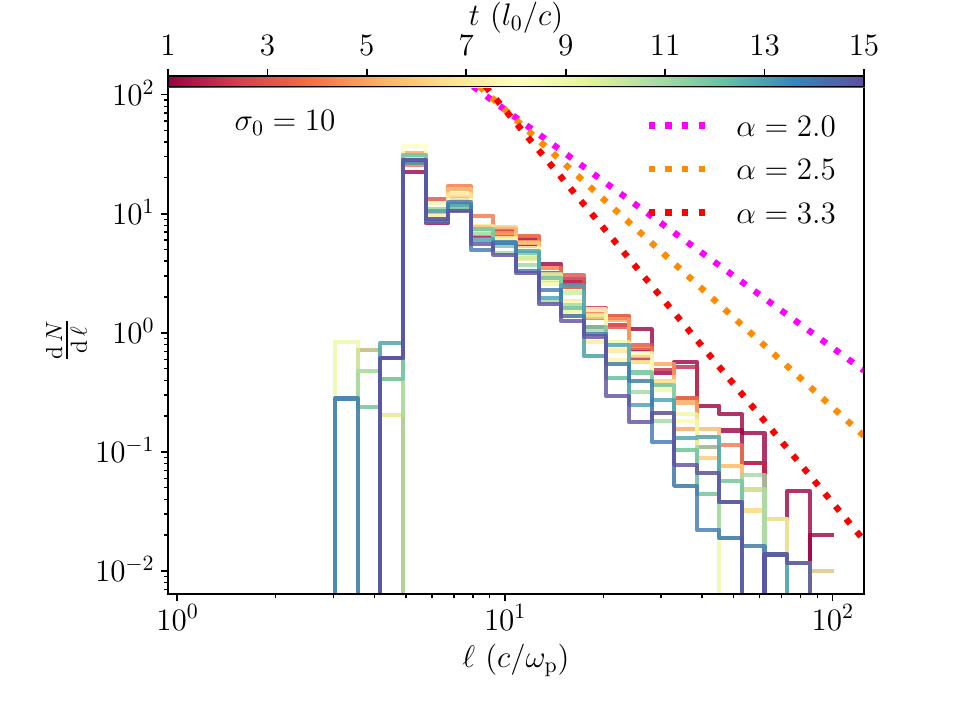}
    \caption{Length distribution, $dN/d\ell$, for simulations with initial $\sigma_0 = (1, 10)$. Three reference power laws with the corresponding index are shown in the legend.}
    \label{fig: ell dist}
    \includegraphics[width=0.45\columnwidth]{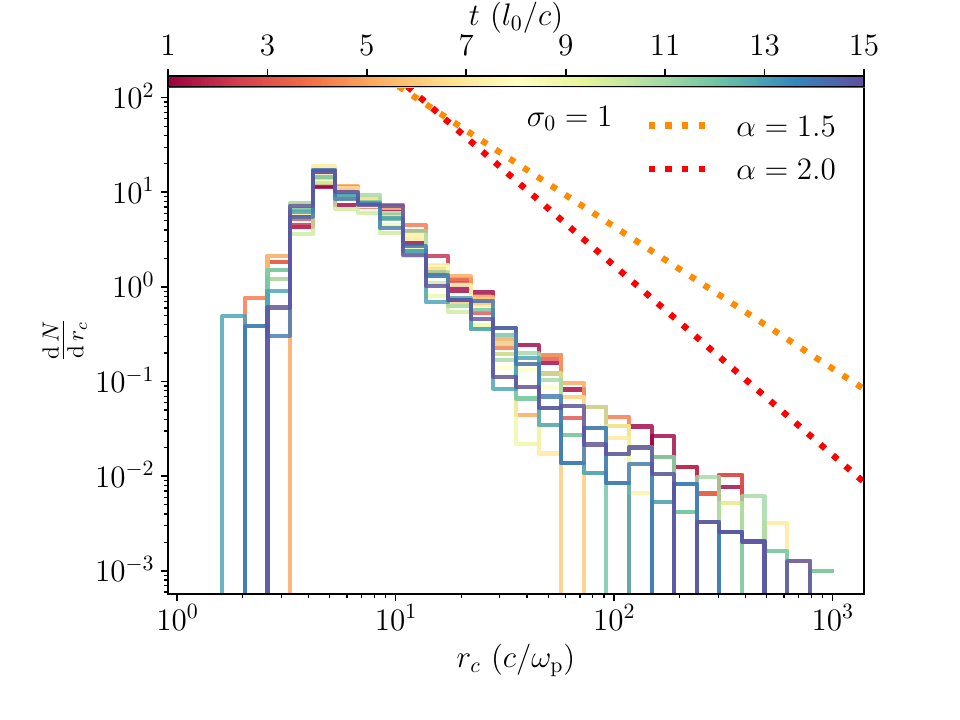}
    \includegraphics[width=0.45\columnwidth]{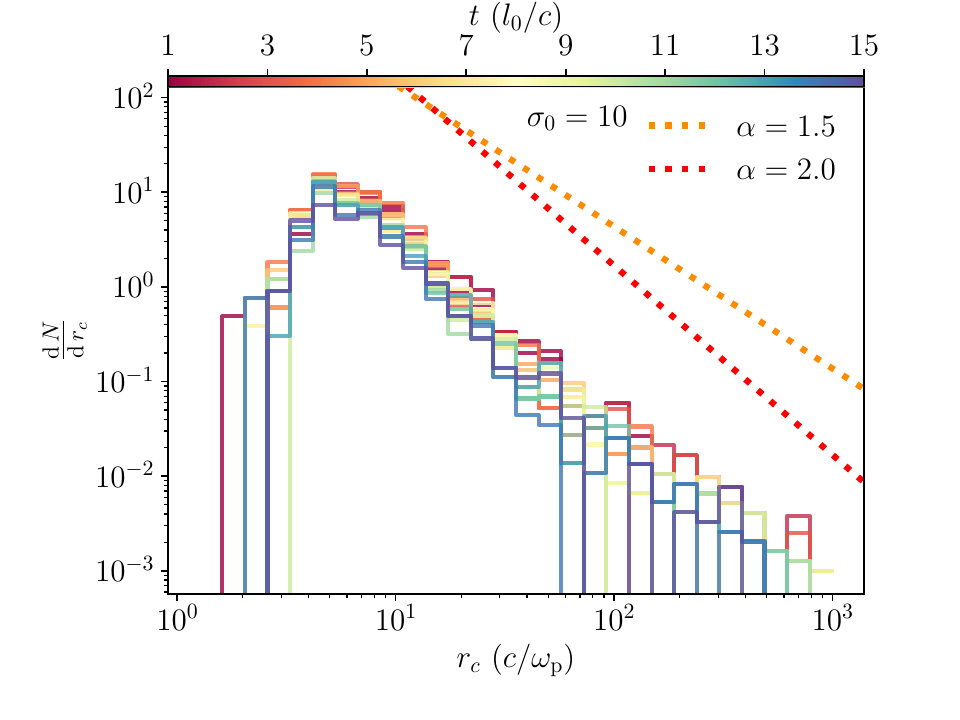}
    \caption{Radius of curvature distributions, $dN/dr_{c}$, for simulations with initial $\sigma_0 = (1, 10)$. Two reference power laws with the corresponding index are shown in the legend. This distribution is restricted to current sheets with $\ell > 7.5 c/\omega_p$, to avoid spurious measurements from small structures.}
    \label{fig: rc dist}
\end{figure*}

\begin{figure*}
\centering
    \includegraphics[width=0.45\columnwidth]{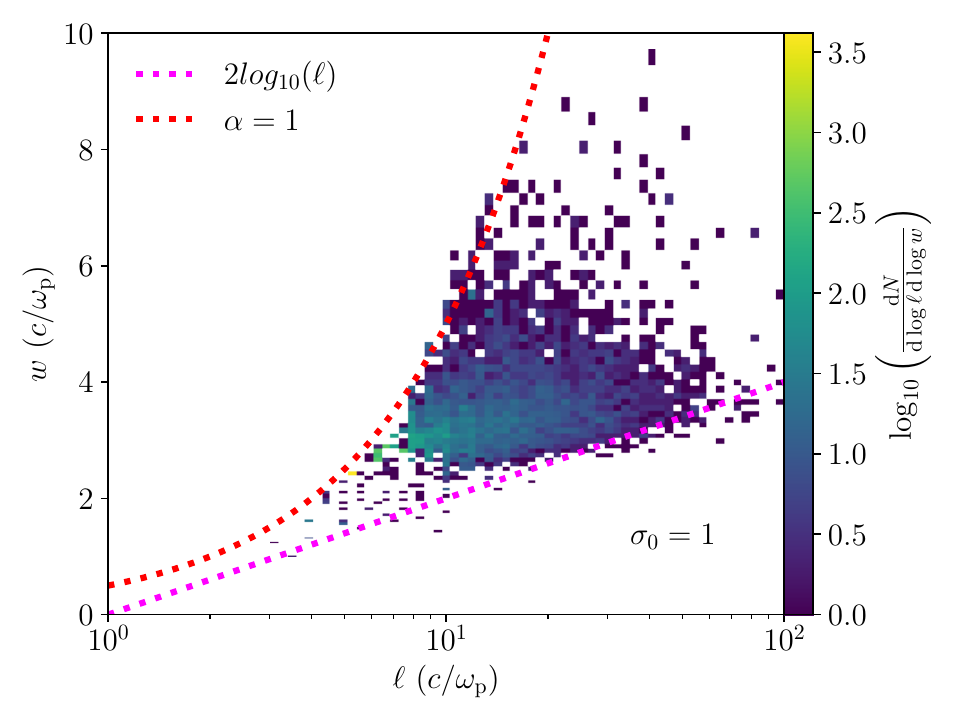}
    \includegraphics[width=0.45\columnwidth]{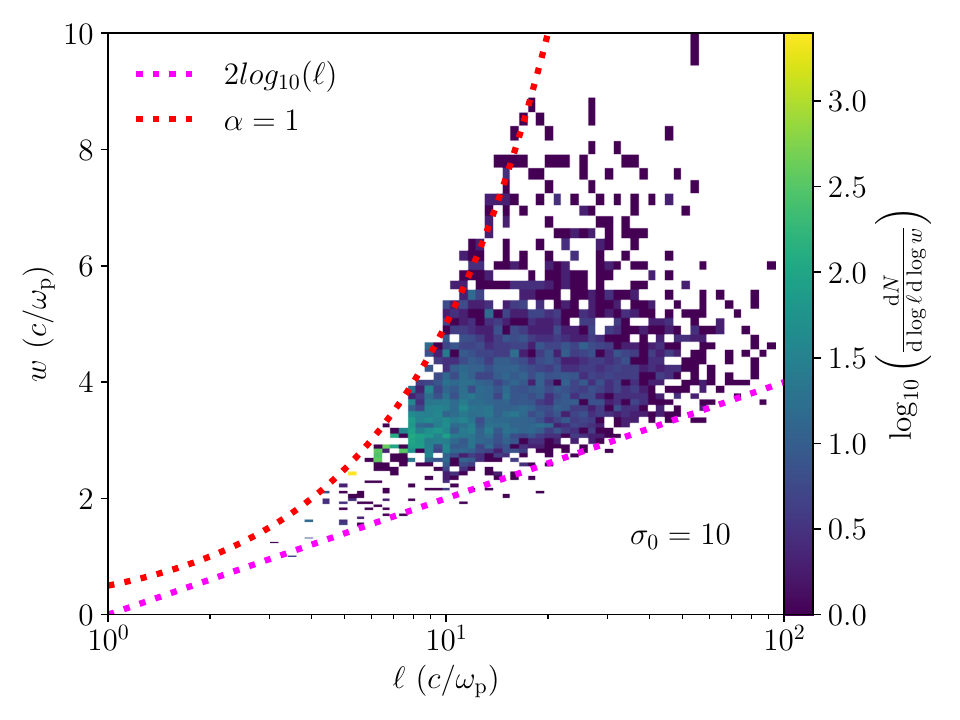}
    \caption{2D histogram correlating the average width and length distributions for simulations with initial $\sigma_0 = 1, 10$. The power law index fits the distributions shown in the legend.}
    \label{fig: 2d_hist_wvsell}
    \includegraphics[width=0.45\columnwidth]{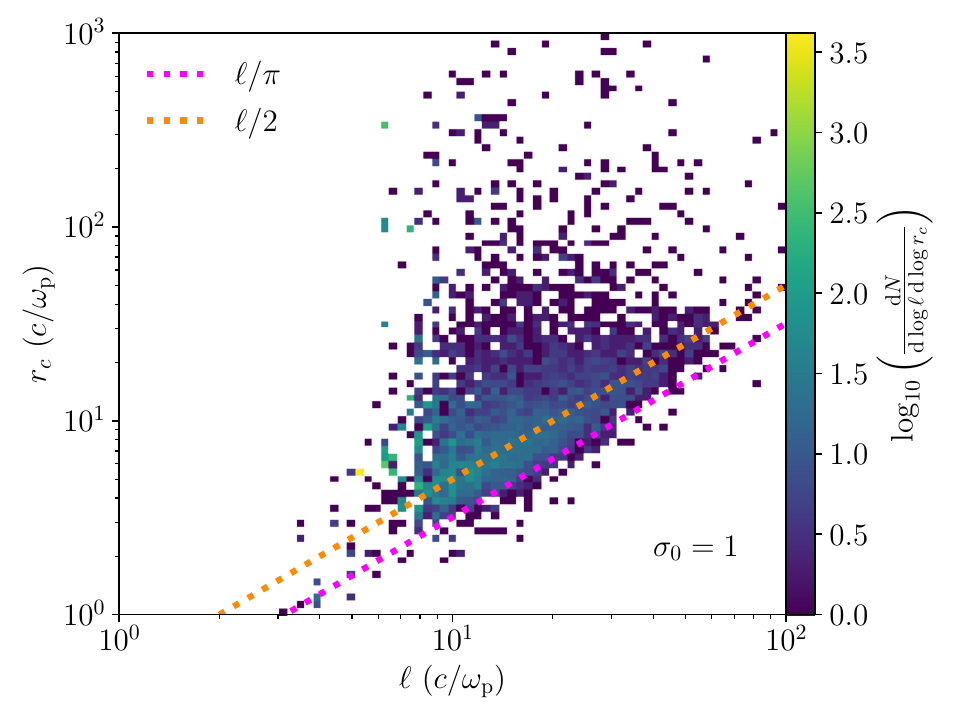}
    \includegraphics[width=0.45\columnwidth]{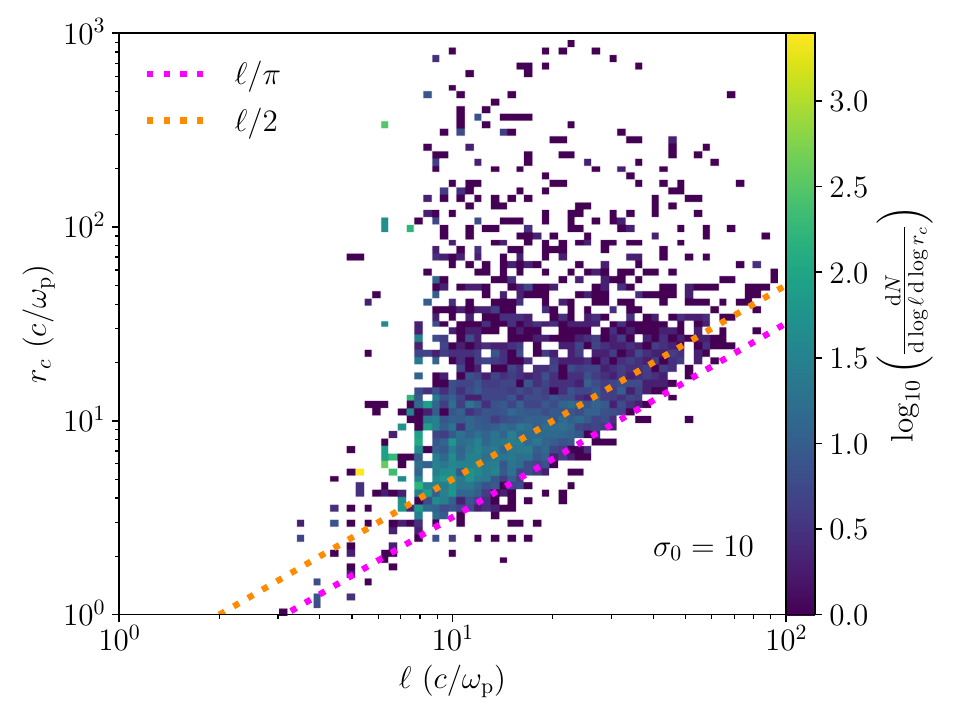}
    \caption{2D histogram correlating the radius of curvature and length distributions for simulations with initial $\sigma_0 = 1, 10$. The power law index fits the distributions shown in the legend. }
    \label{fig: 2d_hist_rcvsell}
    \includegraphics[width=0.45\columnwidth]{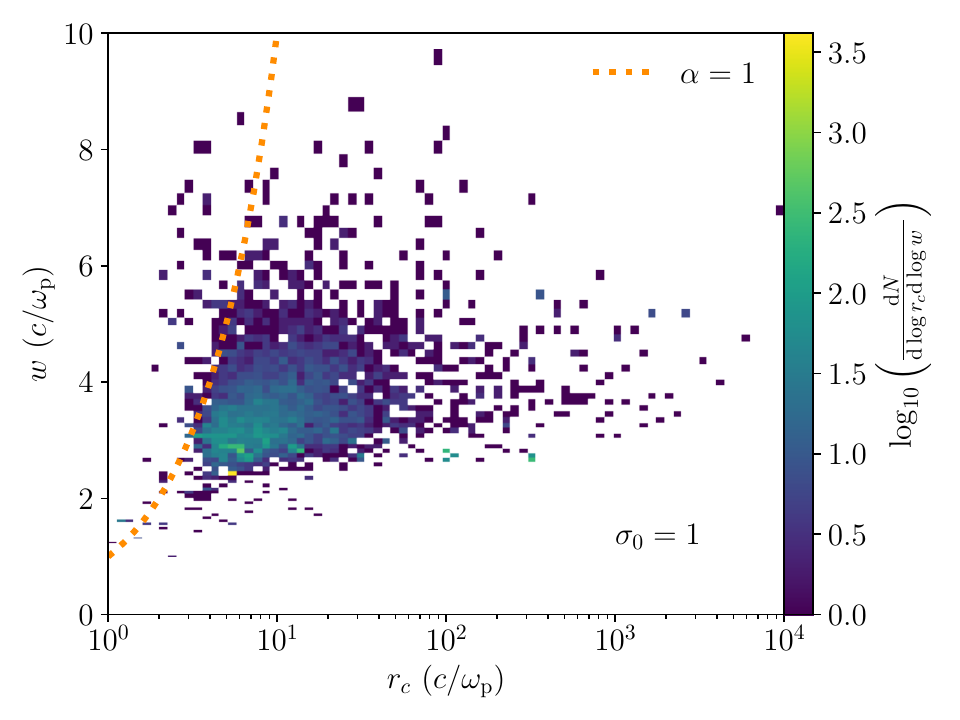}
    \includegraphics[width=0.45\columnwidth]{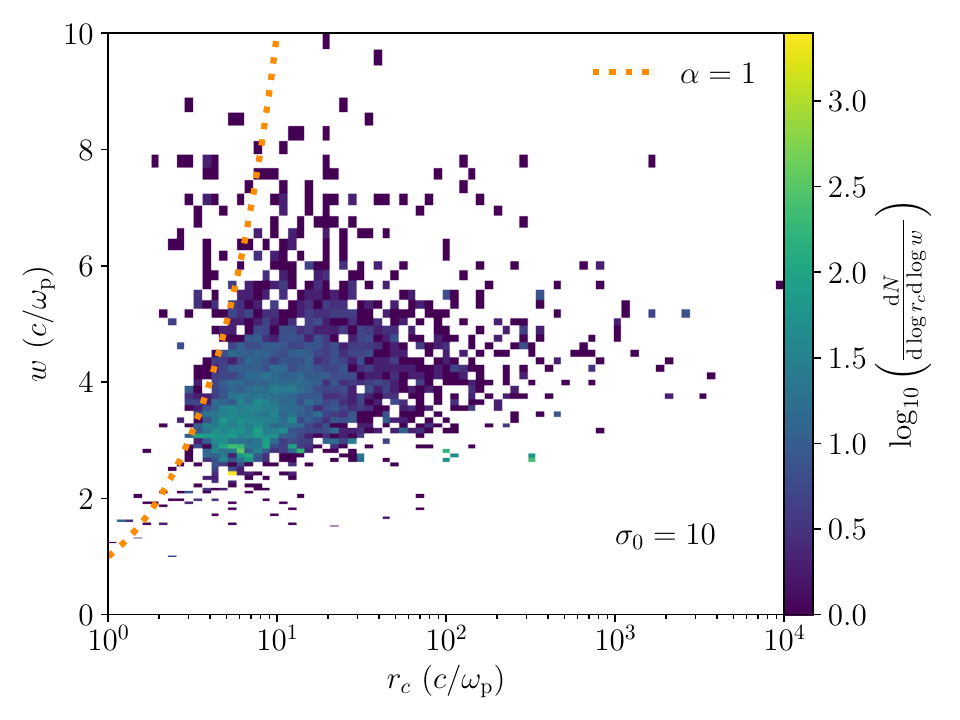}
    \caption{2D histogram correlating the width and radius of curvature distributions for simulations with initial $\sigma_0 = 1, 10$. The power law index fits the distributions shown in the legend.}
    \label{fig: 2d_hist_wvsrc}
\end{figure*}

In Fig.~\ref{fig: width dist}, we plot the width distribution, $P_w(w)$, of our current sheets for both values of $\sigma_0 = 1$ and $10$. 
The distribution resembles an exponential distribution, peaking at around $3 \,  c/\omega_{\mathrm{p}}$. 
Qualitatively, the distributions are best modeled with an exponential function, $\exp(-w\omega_{\mathrm{p}}/c)$.
The width distribution does not greatly vary over time.  

In Fig.~\ref{fig: ell dist}, we visualize the length distribution of our current sheets, $P_{\ell}(\ell)$, and similarly to the width distribution, we find that the distribution remains roughly constant over time. 
The distribution can be modeled as a power-law distribution with an index, $\alpha = 2.5$. 
The distribution peaks at $5 c/\omega_{\mathrm{p}}$, and extends to $\sim\!100 \, c/\omega_{\mathrm{p}}$; i.e., close to the initial energy-carrying scale $l_0$ of the simulation.

Next, in Fig.~\ref{fig: rc dist}, we visualize the curvature distribution $P_{r_c}(r_c)$ (although we omit structures with $\ell < 7.5 c/\omega_{\mathrm{p}}$ because the curvature measurement becomes unreliable for very short structures. 
The distribution is a clear power-law with an index $\alpha = 2.0$. 
The distribution peaks at $3 \, c/\omega_{\mathrm{p}}$, and extends to about $1000 \, c/\omega_\mathrm{p}$. 
The distribution is steady in time, with no significant evolution. 

Furthermore, we study the 2D joint distributions of length and width, $P_{\ell,w}(\ell,w)$, in Fig.~\ref{fig: 2d_hist_wvsell}. 
We find no clear scaling between the two quantities. 
The distribution is bounded such that the upper limits of the distribution are fit by $w \propto \ell^{\alpha}$, where $\alpha = 1.0$. We fit $2\log_{10}(\ell)$ for the lower bound.

Similarly, in Fig.~\ref{fig: 2d_hist_rcvsell}, we visualize the 2D distributions of length and curvature, $P_{\ell,r_c}(\ell,r_c)$. 
The mean value of the sheets linearly scales with $\ell$ (dark orange line). This suggests that the relative curvature to size is the same for all $\ell$.  
Additionally, we find that the distribution can be traced with $r_{c} \propto \ell/\pi$. 
We also show a fit that corresponds to $r_{c} \propto \ell^{\alpha}/2$, where $\alpha = 1$. One final note is that at $\ell < 7 c/\omega_{\mathrm{p}}$ the curvature measurement becomes unreliable and one can see small ``cluster" of points at $r_c \approx 100 c/\omega_{\mathrm{p}}$ and $r_c \approx 300c/\omega_{\mathrm{p}} - 400 c/\omega_{\mathrm{p}}$.   

Finally, in Fig.~\ref{fig: 2d_hist_wvsrc}, we visualize the 2D distribution of the width and curvature, $P_{r_c,w}(r_c,w)$. 
Here, no strong correlation is seen. 
The upper end of the distribution is traced with $w \propto \ell^{\alpha}$, where $\alpha = 1$. Since the structures have widths close to the dissipation scale, it seems reasonable that they are independent of the curvature, which is more likely sensitive to MHD-scale fluctuations.

\section{Discussion} \label{sect:Discussion}

We have analyzed various length statistics of tens of thousands of current sheets found in freely evolving (decaying) plasma turbulence.
In particular, we have found that:
\begin{itemize}
        \item The current sheet length distribution has a power law tail with index $\alpha = 3.3$; it is shallower ($\alpha \approx 2$) at small lengths ($\ell \lesssim 30 \, c/\omega_{\mathrm{p}}$).
        \item The width distribution is sharply peaked at $w \sim 3\, c/\omega_{\mathrm{p}}$; it resembles an exponential decay at higher $w$. There is a slight logarithmic increase of $w$ as a function of $\ell$.
        \item The characteristic curvature radius is $r_c \sim \ell/2$, hence the sheets are slightly curved. However, we find that short ($\ell \sim 10\, c/\omega_{\mathrm{p}}$) sheets are often flatter (larger $r_{c}$); hence, the curvature distribution is a shallower distribution than the length distribution (as seen by the 1D histogram indices), with $\alpha = 2.0$. 
        \item All distributions have weak (or no) time dependence as the turbulence evolves during the studied period of fully developed turbulence.
        \item All distributions have weak (or no) magnetization dependency in the considered range $1 \leq \sigma_0 \leq 10$.
        \item In addition, we calibrated the machine-learning-based identification algorithm of \citet{2021SPIC...9916450B} against the thresholding method. We found that the filling fraction of current sheets in our algorithm matches the thresholding method at a value of $J_{\mathrm{thr}}/J_{\mathrm{rms}} = 3$. 

\end{itemize}

\noindent
The limitations of our study are mainly due to our algorithm being 2D. 
This is a simplification since, in this case, the sheets have a simpler topology to characterize.
In 3D, the current sheets can have much more complex shapes. 
Ultimately, however, the analysis must be generalized to full 3D to obtain physically realistic sheet distributions.

Many possible extensions of the work can also be devised. 
For example, it is possible to measure plasma parameters (e.g., plasma-$\beta$, real magnetization $\sigma$, temperature $T$, the ratio of reversing field to guide field $\delta B/B_{0}$, etc.) upstream from the current sheet using the SEC algorithm, as done in \citet{nurisso_etal_2023}, to gather information about the particle acceleration conditions. 
The sheet statistics should also be studied in different parameter regimes by varying, e.g., 
the initial amplitude $\delta B/B$, 
helicity, and
plasma composition (e.g., using an electron-ion rather than pair plasma composition).
In addition, different driving methods should be explored, such as via wave collisions or purely compressive/solenoidal driving. 
These are expected to change the turbulence intermittency and affect the sheet statistics. 
The presented work is, therefore, only the first step in this direction. 

Our findings are relevant to a broad range of astrophysical systems because kinetic/MHD turbulence is also ubiquitous in those systems. 
Some problems where intermittency of kinetic/MHD turbulence plays a crucial role are:
\begin{itemize}
\item Black hole accretion flows and jets in low luminosity active galactic nuclei. 
Their environments are expected to be magnetized and turbulent  \citep[e.g.,][]{nattila2024}. 
Most of the current efforts in supermassive black-hole accretion flow modeling employ
general relativistic magnetohydrodynamic (GRMHD) simulations \citep{EHT2019_5_2019}. 
However, ideal fluid simulations cannot self-consistently model the temperature of electrons so the electron temperature profile is often ``painted on'' via post-processing (e.g., \citealt{Moscibrodkzka2016}; however, see \citealt{Galishnikova2023} for a recent fully kinetic GRPIC simulations with realistic electron plasma physics).
Alternatively, some GRMHD simulations employ a variety of sub-grid models to compute the electron and ion temperatures based on local PIC simulations of magnetic reconnection \citep{Ball2018, Werner2018} or kinetic/MHD turbulent heating based models \citep{Howes2010, Kawazura2019, meringolo_etal_2023}.

\item The Crab Pulsar Wind Nebula (PWN) is another example of a highly magnetized $(\sigma_0 \gtrsim 1)$ environment that can accelerate particles to very high energies \citep{Rees&Gunn74, Kennel84a, Kennel84b, Atoyan99}. 
Recently, in an attempt to resolve observational discrepancies of the Crab PWN with theoretical models, \citet{Luo2020} proposed a turbulent origin of the non-thermal emission in the Crab PWN. 

\item Cosmic ray diffusion in galactic environments, such as interstellar and inter-cluster mediums, strongly depends on the turbulence physics \citep{Ruszkowski2023}. 
It is still not understood what exactly scatters the cosmic rays. 
However, one hypothesis is the intermittent structures in the turbulence \citep{Kempski2023, Lemoine2023}.
\end{itemize}

\noindent
In summary, we have obtained the scaling distributions for the geometric properties of current sheets, including the length, width, and radius of curvature. 
We find that the length and radius of curvature distributions are well-approximated as power-law distributions and the width distribution with an exponential function. 
This information can be used to develop realistic subgrid models for turbulent heating in the accretion flows of compact objects and other turbulent environments.

\section*{Acknowledgments}
The authors thank Lorenzo  Sironi and Trung Ha for their helpful discussions. 
RS is also thankful to Stack Overflow user (Scott), and the Flatiron Institute for Guest Researcher appointment.  
This work is supported by an ERC grant (ILLUMINATOR, 101114623). 
V.Z. acknowledges support from NSF grant PHY-2409316.
J.N. and V.Z. acknowledge support from a Flatiron Research Fellowship at the Flatiron Institute. 
Research at the Flatiron Institute is supported by the Simons Foundation.

\newpage
\appendix

\section{Transformation to local coordinate frame}
\label{app:1}
In this Appendix, we describe the local coordinate system constructed for presenting example current sheets in Fig.~\ref{fig: sheet gallery} and for measuring the width. 
This coordinate system is defined individually for each structure. 

For each current sheet, we first need the perimeter points of the polygon, $(x_i, y_i)$ with $i \in \{ 1, \dots , N_p \}$, where $N_p$ is the number of perimeter points identified by the SOM algorithm.%
\footnote{
Here, two special cases need to be discussed: 
i) what happens if multiple points satisfy the criterion for being the maximum or minimum coordinates? 
This tends to be the case for structures with complex topologies (e.g. if the structure is a branch-like shape or deformed). 
We checked that these include at most $\approx 60$ sheets out of our sample. Thus, they are statistically negligible, and we will ignore them. 
ii) How to account for the splitting of structures when they cross the periodic boundary of the simulation? 
For simplicity, we do not account for this;
However, in that case, it should be noted that these structures are double-counted. 
We verified that there are $\mathcal{O}(10)$ such sheets per time step, so they are not expected to influence our analysis.
}
First, we identify the left (or right) edges of the polygon by finding the index $ i_{\rm L}$ (or $i_{\rm R}$) at which $x_i$ takes a minimal (or maximal) value in the set $\{ x_i \}$; we denote the (Euclidean) distance between these points by $r_x = [(x_{i_{\rm R}}-x_{i_{\rm L}})^2 + (y_{i_{\rm R}}-y_{i_{\rm L}})^2]^{1/2}$. 
Likewise, we identify the bottom (or top) edge of the polygon by finding the index $i_{\rm B}$ (or $i_{\rm T}$) at which $y_i$ takes a minimal (or maximal) value in the set $\{ y_i \}$; we denote the distance between these points by $r_y = [(x_{i_{\rm T}}-x_{i_{\rm B}})^2+(y_{i_{\rm T}}-y_{i_{\rm B}})^2]^{1/2}$. 
The ``major axis'' is then defined by the larger of $r_x$ and $r_y$.%
\footnote{Note that this procedure is not coordinate invariant since it treats separations along the $x$ and $y$ directions as special.}

Next, we transform to a new coordinate system $(x',y')$ such that the center of the polygon and the major axis is along the $x'$-axis. 
To achieve this, we define a vector $\boldsymbol{D}$ that is oriented along the major axis, 
\[   
\boldsymbol{D} = 
     \begin{cases}
       \frac{(x_{i_{\rm R}} - x_{i_{\rm L}}, y_{i_{\rm R}} - y_{i_{\rm L}})}{r_x} \, &  \, r_x > r_y \\ 
       \frac{(x_{i_{\rm T}} - x_{i_{\rm B}}, y_{i_{\rm T}} - y_{i_{\rm B}})}{r_y} \, &  \, \mathrm{otherwise}\\ 
     \end{cases}
\]
and a rotation matrix
\begin{equation}
    \label{eqn: rotation matrix}
    M = 
    \begin{bmatrix}
        D_x & D_y \\
        -D_y & D_x \\ 
    \end{bmatrix}
    \, .
\end{equation}
To transform to the new coordinates $(x',y')$ for each structure, we multiply the original coordinates $(x,y)$ by the rotation matrix $M$ and then subtract the structure's ``average" location from it's position coordinates: 
\begin{equation}
\label{eq: primecoordinates}
   (x'_i,y'_i) = M \cdot (x_i,y_i) - (\langle x_i \rangle, \langle y_i \rangle) \,,
\end{equation}
where $\langle x_i \rangle$, and $\langle y_i \rangle$ are the mean location values from the sets ${x_i}$, and $y_{i}$, respectively. 

\section{Algorithm for measuring the sheet width}
\label{app:2}

Continuing the calculation from Appendix ~\ref{app:1}, we are now in the local coordinate frame $(x'_i, y'_i)$. 
In this frame, we calculate the width of each current sheet. 
First, we must re-identify the left (or right) edges of the polygon in the local coordinate frame; $i'_{\rm L}$ (or $i'_{\rm R}$) at which $x'_i$ takes a minimal (or maximal) value in the set $\{ x'_i \}$ of points on the structure's perimeter. Likewise, we re-identify the bottom (or top) edge of the polygon by finding the index $i'_{\rm B}$ (or $i'_{\rm T}$) at which $y'_i$ takes a minimal (or maximal) value in the set $\{ y'_i \}$. 

One simple method to calculate the width would be to take the (Euclidean) distance between the points $(x'_{i'_{\rm T}}, y'_{i'_{\rm T}})$ and $(x'_{i'_{\rm B}}, y'_{i'_{\rm B}})$. 
However, this definition may not represent the overall structure if the current sheet's width varies along its length. 

Therefore, to define a better representation of the overall width, we take the additional step of breaking each current sheet into vertical strips and averaging the width across the strips. 
The first step is to divide each sheet into a ``top" half and ``bottom" half along the direction parallel to $x'_i$. 
We begin from $x'_{i'_{\rm L}}$, and find its $y'$-coordinate $y'_{i'_{\rm L}}$, and similarly, we take $x'_{i'_{\rm R}}$ and find it's $y'$-coordinate pair $y'_{i'_{\rm R}}$.  
Next, we begin sorting the $y'$-coordinate pairs into two new arrays, $y_{\mathrm{i, top}}$ and $y_{\mathrm{i, bot}}$ in the following manner: We begin from $x'_{i}$ at $i=1$ and begin looping through $i$ until $x'_{i'_{\rm L}} = x'_{i}$, or $x'_{i'_{\rm R}} = x'_{i}$. In the former case, we append the values of $y'_{i}$ to $y_{\mathrm{i, top}}(x'_{i})$, and in the later case we append $y'_{i}$ to $y_{\mathrm{i, bot}}(x'_{i})$. This process iterates until we reach $i = N_{p}$. 

Next, we use the \texttt{scipy.interpolate} subpackage \citep{2020SciPy-NMeth} to interpolate between $y'_{\mathrm{i, top}}(x'_{i})$ and $y'_{\mathrm{i, bot}}(x'_{i})$ using an array of points linearly spaced between $x'_{i'_{\rm L}}$ and $x'_{i'_{\rm R}}$, thus creating two new arrays $y''_{\mathrm{top}}(x'_{i})$ and $y''_{\mathrm{bot}}(x'_{i})$ of the $y'$ coordinates corresponding to $\{x'_i\}$. 
This allows us to create vertical slices (along the $y'$-axis), and then we average over the sum of the slices to obtain our width, 

\begin{equation}
\label{eqn: width}
w = \frac{1}{N_x} \sum_{i = 1}^{N_p} \left|\, y''_\mathrm{top}(x'_{i}) - y''_\mathrm{bot}(x'_{i}) \,\right| \,
\end{equation}
where $N_x = 50$ is the number of points we use to generate the interpolator. We have found this method to yield a more robust estimate of $w$ in comparison to just using a slice (e.g., $w \approx |y''_\mathrm{top}(x'_s) - y''_\mathrm{bot}(x'_s)|$ at location $x'_s = \frac{1}{2}| x'_{i_{\rm L}} - x'_{i_{\rm R}}|$) or averaging using the number of available pixels only. The reason for that is that the width of the sheet can vary along the length.

\section{Algorithm for calculating the parameters of a circle fitted through 3 points}
\label{app:3}

Here, we summarize the algorithm that we use to calculate the radius of curvature of a given current sheet and the center of the circle. 
The algorithm is based on the curve fitting algorithms listed in \citet{Umbach2003}. 
More sophisticated methods, using more than three points, could also be used to obtain a ``best fit" circle instead.

We begin by choosing three points in our current sheet; the first two are $x'_{i_{L}}$ and $x'_{i_{R}}$ described above, and the third point is obtained by first calculating the half width at the center of the sheet using (\ref{eqn: width}); i.e., $ y_{h} = 1/2|y''_\mathrm{top}(x'_s) - y''_\mathrm{bot}(x'_s)|$ at location $x'_s = \frac{1}{2}| x'_{i_{\rm L}} - x'_{i_{\rm R}}|$. Then we obtain the midpoint $y_{m} = y_{h} + y''_\mathrm{bot}(x'_s)$. These points are marked by red dots in Fig.~\ref{fig: sheet gallery}. Next we map the  three points into the complex numbers $(z'_{1}, z'_{2}, z'_{3})$; using the definition $z' = x' + \sqrt{-1} y'$. 

We then apply a linear transformation to $(z'_{1}, z'_{2}, z'_{3})$; i.e 
\begin{equation}
    z'_{j} \rightarrow \frac{z'_{j} - z'_{1}}{z'_{2}- z'_{1}}\,
\end{equation}
which results in
$z'_1 \rightarrow 0$, $z'_2 \rightarrow 1$, and  $z'_3 \rightarrow (z'_{3} - z'_{1})/(z'_{2}- z'_{1}) \equiv u$. $z'_1$ is the left point of the structure, $z'_2$ is the right point of the structure, and finally, $z'_3$ is the middle point in the structure. 

We can then use the relationship $|z'_{j} - C| = r_{c}$ (where $r_{c}$ is the radius of curvature of the circle fit, $C$ is the coordinates of the circle's center, and $j \in \{ 1,2,3 \})$ to obtain the following system of equations: 
\begin{equation}
    \label{eqn: complex z1}
 |z'_1 - C|^2 = |C|^2 = r_{c}^{2} \, , \\
\end{equation}
\begin{equation}
    \label{eqn: complex z2}
 |z'_2 - C|^2 =  1 - C - \overline{C} + |C^2| = r_{c}^{2} \, , \\
\end{equation}
\begin{equation}
    \label{eqn: complex z3}
   |z'_3 - C|^2 = |u|^2 - \overline{u}C - u\overline{C} + |C|^2 = r_{c}^2 \, .
\end{equation}
We solve $C$ from Eqs (\ref{eqn: complex z1})-(\ref{eqn: complex z3}). The center of the circle is then, 
\begin{equation}
    \label{eqn: complex circle center linear}
    C = \frac{u - |u|^2}{u - \overline{u}} \, .
\end{equation}
Finally, we undo the linear transformation, which yields the center of the circle in our original coordinates, 
\begin{equation}
    \label{eqn: complex center org}
    C \rightarrow (z'_{2} - z'_{1})\frac{u - |u|^2}{u - \overline{u}} + z'_{1} \, .
\end{equation}
Therefore, we now have $C$, and using Eq.~(\ref{eqn: complex z1}), we obtain the radius of curvature, $r_{c}$.

\bibliographystyle{jpp}
\bibliography{refs}

\begin{thebibliography}{14}
\expandafter\ifx\csname natexlab\endcsname\relax\def\natexlab#1{#1}\fi
\def\au#1{#1} \def\ed#1{#1} \def\yr#1{#1}\def\at#1{#1}\def\jt#1{\textit{#1}}
  \def\bt#1{#1}\def\bvol#1{\textbf{#1}} \def\vol#1{#1} \def\pg#1{#1}
  \def\publ#1{#1}\def\arxiv#1{#1}\def\org#1{#1}\def\st#1{\textit{#1}}

\bibitem[Batchelor(1971)]{Batchelor59}
{\sc \au{Batchelor, G.~K.}} \yr{1971}  \at{Small-scale variation of convected
  quantities like temperature in turbulent fluid. part 1. general discussion
  and the case of small conductivity.}  \jt{J.~Fluid Mech.}  \bvol{5},
  \pg{113--133}.

\bibitem[Brownell \& Su(2004)]{Brownell04}
{\sc \au{Brownell, C.~J.} \& \au{Su, L.~K.}} \yr{2004}  \at{Planar measurements
  of differential diffusion in turbulent jets}.  \jt{AIAA Paper 2004-2335} .

\bibitem[Brownell \& Su(2007)]{Brownell07}
{\sc \au{Brownell, C.~J.} \& \au{Su, L.~K.}} \yr{2007}  \at{Scale relations and
  spatial spectra in a differentially diffusing jet}.  \jt{AIAA Paper
  2007-1314} .

\bibitem[Dennis(1985)]{Dennis85}
{\sc \au{Dennis, S. C.~R.}} \yr{1985}  \at{{Compact explicit finite difference
  approximations to the Navier--Stokes equation}}.  \bt{In {\em Ninth Intl
  Conf. on Numerical Methods in Fluid Dynamics\/} (ed. \ed{Soubbaramayer \&
  J.~P. Boujot})},  \st{Lecture Notes in Physics},  \vol{vol. 218},  \pg{pp.
  23--51}.  \publ{Springer}.

\bibitem[Galtier {\em et~al.\/}({2000})Galtier, Nazarenko, Newell \&
  Pouquet]{Galtier00}
{\sc \au{Galtier, S.}, \au{Nazarenko, S.}, \au{Newell, A.} \& \au{Pouquet, A.}}
  \yr{{2000}}  \at{{A weak turbulence theory for incompressible
  magnetohydrodynamics}}.  \jt{{J.~Plasma Phys.}}  \bvol{{63}}~({5}),
  \pg{{447--488}}.

\bibitem[Koch(1983)]{Koch83}
{\sc \au{Koch, W.}} \yr{1983}  \at{Resonant acoustic frequencies of flat plate
  cascades}.  \jt{J.~Sound Vib.}  \bvol{88},  \pg{233--242}.

\bibitem[Lee(1971)]{Lee71}
{\sc \au{Lee, J.-J.}} \yr{1971}  \at{Wave-induced oscillations in harbours of
  arbitrary geometry}.  \jt{J.~Fluid Mech.}  \bvol{45},  \pg{375--394}.

\bibitem[Linton \& Evans(1992)]{Linton92}
{\sc \au{Linton, C.~M.} \& \au{Evans, D.~V.}} \yr{1992}  \at{The radiation and
  scattering of surface waves by a vertical circular cylinder in a channel}.
  \jt{Phil.\ Trans.\ R. Soc.\ Lond.}  \bvol{338},  \pg{325--357}.

\bibitem[Martin(1980)]{Martin80}
{\sc \au{Martin, P.~A.}} \yr{1980}  \at{On the null-field equations for the
  exterior problems of acoustics}.  \jt{Q.~J. Mech.\ Appl.\ Maths}  \bvol{33},
  \pg{385--396}.

\bibitem[Miller(1991)]{Miller91}
{\sc \au{Miller, P.~L.}} \yr{1991}  \at{Mixing in high schmidt number turbulent
  jets}. PhD thesis, California Institute of Technology.

\bibitem[Rogallo(1981)]{Rogallo81}
{\sc \au{Rogallo, R.~S.}} \yr{1981}  \bt{Numerical experiments in homogeneous
  turbulence}. {\em Tech. Rep.\/} 81835.  \org{NASA Tech.\ Mem.}

\bibitem[Ursell(1950)]{Ursell50}
{\sc \au{Ursell, F.}} \yr{1950}  \at{Surface waves on deep water in the
  presence of a submerged cylinder i}.  \jt{Proc.\ Camb.\ Phil.\ Soc.}
  \bvol{46},  \pg{141--152}.

\bibitem[{van Wijngaarden}(1968)]{Wijngaarden68}
{\sc \au{{van Wijngaarden}, L.}} \yr{1968}  \at{On the oscillations near and at
  resonance in open pipes}.  \jt{J.~Engng Maths}  \bvol{2},  \pg{225--240}.

\bibitem[Worster(1992)]{Worster92}
{\sc \au{Worster, M.~G.}} \yr{1992}  \at{{The dynamics of mushy layers}}.
  \bt{In {\em In Interactive dynamics of convection and solidification\/} (ed.
  \ed{S.~H. Davis, H.~E. Huppert, W.~Muller \& M.~G. Worster})},  \pg{pp.
  113--138}.  \publ{Kluwer}.

\end{thebibliography}


\begin{thebibliography}{59}
\expandafter\ifx\csname natexlab\endcsname\relax\def\natexlab#1{#1}\fi
\def\au#1{#1} \def\ed#1{#1} \def\yr#1{#1}\def\at#1{#1}\def\jt#1{\textit{#1}}
  \def\bt#1{#1}\def\bvol#1{\textbf{#1}} \def\vol#1{#1} \def\pg#1{#1}
  \def\publ#1{#1}\def\arxiv#1{#1}\def\org#1{#1}\def\st#1{\textit{#1}}

\bibitem[{Atoyan}(1999)]{Atoyan99}
{\sc \au{{Atoyan}, A.~M.}} \yr{1999}  \at{{Radio spectrum of the Crab nebula as
  an evidence for fast initial spin of its pulsar}}.  \jt{\aap}  \bvol{346},
  \pg{L49--L52},  \arxiv{arXiv: astro-ph/9905204}.

\bibitem[Azizabadi {\em et~al.\/}(2021)Azizabadi, Jain \&
  B{\"u}chner]{azizabadi_etal_2021}
{\sc \au{Azizabadi, A.~C.}, \au{Jain, N.} \& \au{B{\"u}chner, J.}} \yr{2021}
  \at{Identification and characterization of current sheets in collisionless
  plasma turbulence}.  \jt{Physics of Plasmas}  \bvol{28}~(5).

\bibitem[{Ball} {\em et~al.\/}(2018){Ball}, {Sironi} \& {{\"O}zel}]{Ball2018}
{\sc \au{{Ball}, D.}, \au{{Sironi}, L.} \& \au{{{\"O}zel}, F.}} \yr{2018}
  \at{{Electron and Proton Acceleration in Trans-relativistic Magnetic
  Reconnection: Dependence on Plasma Beta and Magnetization}}.  \jt{\apj}
  \bvol{862}~(1),  \pg{80},  \arxiv{arXiv: 1803.05556}.

\bibitem[Beresnyak(2019)]{Beresnyak2019}
{\sc \au{Beresnyak, A.}} \yr{2019}  \at{Mhd turbulence}.  \jt{Living Reviews in
  Computational Astrophysics}  \bvol{5}~(1).

\bibitem[{Biskamp}(2008)]{2008matu.book.....B}
{\sc \au{{Biskamp}, D.}} \yr{2008} {\em {Magnetohydrodynamic Turbulence}\/}.

\bibitem[Borgogno {\em et~al.\/}(2022)Borgogno, Grasso, Achilli, Rom{\'e} \&
  Comisso]{borgogno_etal_2022}
{\sc \au{Borgogno, D.}, \au{Grasso, D.}, \au{Achilli, B.}, \au{Rom{\'e}, M.} \&
  \au{Comisso, L.}} \yr{2022}  \at{Coexistence of plasmoid and
  kelvin--helmholtz instabilities in collisionless plasma turbulence}.  \jt{The
  Astrophysical Journal}  \bvol{929}~(1),  \pg{62}.

\bibitem[{Bussov} \& {N{\"a}ttil{\"a}}(2021)]{2021SPIC...9916450B}
{\sc \au{{Bussov}, M.} \& \au{{N{\"a}ttil{\"a}}, J.}} \yr{2021}
  \at{{Segmentation of turbulent computational fluid dynamics simulations with
  unsupervised ensemble learning}}.  \jt{Signal Processing: Image
  Communication}  \bvol{99},  \pg{116450},  \arxiv{arXiv: 2109.01381}.

\bibitem[{Chernoglazov} {\em et~al.\/}(2023){Chernoglazov}, {Hakobyan} \&
  {Philippov}]{Chernoglazov2023}
{\sc \au{{Chernoglazov}, A.}, \au{{Hakobyan}, H.} \& \au{{Philippov}, A.}}
  \yr{2023}  \at{{High-energy Radiation and Ion Acceleration in
  Three-dimensional Relativistic Magnetic Reconnection with Strong Synchrotron
  Cooling}}.  \jt{\apj}  \bvol{959}~(2),  \pg{122},  \arxiv{arXiv: 2305.02348}.

\bibitem[{Chernoglazov} {\em et~al.\/}(2021){Chernoglazov}, {Ripperda} \&
  {Philippov}]{Chernoglazov_2021}
{\sc \au{{Chernoglazov}, A.}, \au{{Ripperda}, B.} \& \au{{Philippov}, A.}}
  \yr{2021}  \at{{Dynamic Alignment and Plasmoid Formation in Relativistic
  Magnetohydrodynamic Turbulence}}.  \jt{\apjl}  \bvol{923}~(1),  \pg{L13},
  \arxiv{arXiv: 2111.08188}.

\bibitem[{Comisso} \& {Jiang}(2023)]{Comisso2023}
{\sc \au{{Comisso}, L.} \& \au{{Jiang}, B.}} \yr{2023}  \at{{Pitch-angle
  Anisotropy Imprinted by Relativistic Magnetic Reconnection}}.  \jt{\apj}
  \bvol{959}~(2),  \pg{137},  \arxiv{arXiv: 2310.17560}.

\bibitem[{Comisso} \& {Sironi}(2018)]{comisso&sironi2018}
{\sc \au{{Comisso}, L.} \& \au{{Sironi}, L.}} \yr{2018}  \at{{Particle
  Acceleration in Relativistic Plasma Turbulence}}.  \jt{\prl}
  \bvol{121}~(25),  \pg{255101},  \arxiv{arXiv: 1809.01168}.

\bibitem[{Comisso} \& {Sironi}(2019)]{comisso&sironi2019}
{\sc \au{{Comisso}, L.} \& \au{{Sironi}, L.}} \yr{2019}  \at{{The Interplay of
  Magnetically Dominated Turbulence and Magnetic Reconnection in Producing
  Nonthermal Particles}}.  \jt{\apj}  \bvol{886}~(2),  \pg{122},  \arxiv{arXiv:
  1909.01420}.

\bibitem[{Davis} {\em et~al.\/}(2024){Davis}, {Comisso} \&
  {Giannios}]{Davies2024}
{\sc \au{{Davis}, Z.}, \au{{Comisso}, L.} \& \au{{Giannios}, D.}} \yr{2024}
  \at{{Intermittency and Dissipative Structures Arising from Relativistic
  Magnetized Turbulence}}.  \jt{\apj}  \bvol{964}~(1),  \pg{14},  \arxiv{arXiv:
  2311.08627}.

\bibitem[{Event Horizon Telescope Collaboration} {\em et~al.\/}(2019){Event
  Horizon Telescope Collaboration}, {Akiyama}, {Alberdi}, {Alef}, {Asada},
  {Azulay}, {Baczko}, {Ball}, {Balokovi{\'c}}, {Barrett}, {Bintley},
  {Blackburn}, {Boland}, {Bouman}, {Bower}, {Bremer}, {Brinkerink},
  {Brissenden}, {Britzen}, {Broderick}, {Broguiere}, {Bronzwaer}, {Byun},
  {Carlstrom}, {Chael}, {Chan}, {Chatterjee}, {Chatterjee}, {Chen}, {Chen},
  {Cho}, {Christian}, {Conway}, {Cordes}, {Crew}, {Cui}, {Davelaar}, {De
  Laurentis}, {Deane}, {Dempsey}, {Desvignes}, {Dexter}, {Doeleman}, {Eatough},
  {Falcke}, {Fish}, {Fomalont}, {Fraga-Encinas}, {Friberg}, {Fromm},
  {G{\'o}mez}, {Galison}, {Gammie}, {Garc{\'\i}a}, {Gentaz}, {Georgiev},
  {Goddi}, {Gold}, {Gu}, {Gurwell}, {Hada}, {Hecht}, {Hesper}, {Ho}, {Ho},
  {Honma}, {Huang}, {Huang}, {Hughes}, {Ikeda}, {Inoue}, {Issaoun}, {James},
  {Jannuzi}, {Janssen}, {Jeter}, {Jiang}, {Johnson}, {Jorstad}, {Jung},
  {Karami}, {Karuppusamy}, {Kawashima}, {Keating}, {Kettenis}, {Kim}, {Kim},
  {Kim}, {Kino}, {Koay}, {Koch}, {Koyama}, {Kramer}, {Kramer}, {Krichbaum},
  {Kuo}, {Lauer}, {Lee}, {Li}, {Li}, {Lindqvist}, {Liu}, {Liuzzo}, {Lo},
  {Lobanov}, {Loinard}, {Lonsdale}, {Lu}, {MacDonald}, {Mao}, {Markoff},
  {Marrone}, {Marscher}, {Mart{\'\i}-Vidal}, {Matsushita}, {Matthews},
  {Medeiros}, {Menten}, {Mizuno}, {Mizuno}, {Moran}, {Moriyama},
  {Moscibrodzka}, {Mul{\ensuremath{\ddot{}}}ler}, {Nagai}, {Nagar}, {Nakamura},
  {Narayan}, {Narayanan}, {Natarajan}, {Neri}, {Ni}, {Noutsos}, {Okino},
  {Olivares}, {Oyama}, {{\"O}zel}, {Palumbo}, {Patel}, {Pen}, {Pesce},
  {Pi{\'e}tu}, {Plambeck}, {PopStefanija}, {Porth}, {Prather},
  {Preciado-L{\'o}pez}, {Psaltis}, {Pu}, {Ramakrishnan}, {Rao}, {Rawlings},
  {Raymond}, {Rezzolla}, {Ripperda}, {Roelofs}, {Rogers}, {Ros}, {Rose},
  {Roshanineshat}, {Rottmann}, {Roy}, {Ruszczyk}, {Ryan}, {Rygl},
  {S{\'a}nchez}, {S{\'a}nchez-Arguelles}, {Sasada}, {Savolainen}, {Schloerb},
  {Schuster}, {Shao}, {Shen}, {Small}, {Sohn}, {SooHoo}, {Tazaki}, {Tiede},
  {Tilanus}, {Titus}, {Toma}, {Torne}, {Trent}, {Trippe}, {Tsuda}, {van
  Bemmel}, {van Langevelde}, {van Rossum}, {Wagner}, {Wardle}, {Weintroub},
  {Wex}, {Wharton}, {Wielgus}, {Wong}, {Wu}, {Young}, {Young}, {Younsi},
  {Yuan}, {Yuan}, {Zensus}, {Zhao}, {Zhao}, {Zhu}, {Anczarski}, {Baganoff},
  {Eckart}, {Farah}, {Haggard}, {Meyer-Zhao}, {Michalik}, {Nadolski},
  {Neilsen}, {Nishioka}, {Nowak}, {Pradel}, {Primiani}, {Souccar},
  {Vertatschitsch}, {Yamaguchi} \& {Zhang}]{EHT2019_5_2019}
{\sc \au{{Event Horizon Telescope Collaboration}}, \au{{Akiyama}, K.},
  \au{{Alberdi}, A.}, \au{{Alef}, W.}, \au{{Asada}, K.}, \au{{Azulay}, R.},
  \au{{Baczko}, A.-K.}, \au{{Ball}, D.}, \au{{Balokovi{\'c}}, M.},
  \au{{Barrett}, J.}, \au{{Bintley}, D.}, \au{{Blackburn}, L.}, \au{{Boland},
  W.}, \au{{Bouman}, K.~L.}, \au{{Bower}, G.~C.}, \au{{Bremer}, M.},
  \au{{Brinkerink}, C.~D.}, \au{{Brissenden}, R.}, \au{{Britzen}, S.},
  \au{{Broderick}, A.~E.}, \au{{Broguiere}, D.}, \au{{Bronzwaer}, T.},
  \au{{Byun}, D.-Y.}, \au{{Carlstrom}, J.~E.}, \au{{Chael}, A.}, \au{{Chan},
  C.-k.}, \au{{Chatterjee}, S.}, \au{{Chatterjee}, K.}, \au{{Chen}, M.-T.},
  \au{{Chen}, Y.}, \au{{Cho}, I.}, \au{{Christian}, P.}, \au{{Conway}, J.~E.},
  \au{{Cordes}, J.~M.}, \au{{Crew}, G.~B.}, \au{{Cui}, Y.}, \au{{Davelaar},
  J.}, \au{{De Laurentis}, M.}, \au{{Deane}, R.}, \au{{Dempsey}, J.},
  \au{{Desvignes}, G.}, \au{{Dexter}, J.}, \au{{Doeleman}, S.~S.},
  \au{{Eatough}, R.~P.}, \au{{Falcke}, H.}, \au{{Fish}, V.~L.}, \au{{Fomalont},
  E.}, \au{{Fraga-Encinas}, R.}, \au{{Friberg}, P.}, \au{{Fromm}, C.~M.},
  \au{{G{\'o}mez}, J.~L.}, \au{{Galison}, P.}, \au{{Gammie}, C.~F.},
  \au{{Garc{\'\i}a}, R.}, \au{{Gentaz}, O.}, \au{{Georgiev}, B.}, \au{{Goddi},
  C.}, \au{{Gold}, R.}, \au{{Gu}, M.}, \au{{Gurwell}, M.}, \au{{Hada}, K.},
  \au{{Hecht}, M.~H.}, \au{{Hesper}, R.}, \au{{Ho}, L.~C.}, \au{{Ho}, P.},
  \au{{Honma}, M.}, \au{{Huang}, C.-W.~L.}, \au{{Huang}, L.}, \au{{Hughes},
  D.~H.}, \au{{Ikeda}, S.}, \au{{Inoue}, M.}, \au{{Issaoun}, S.}, \au{{James},
  D.~J.}, \au{{Jannuzi}, B.~T.}, \au{{Janssen}, M.}, \au{{Jeter}, B.},
  \au{{Jiang}, W.}, \au{{Johnson}, M.~D.}, \au{{Jorstad}, S.}, \au{{Jung}, T.},
  \au{{Karami}, M.}, \au{{Karuppusamy}, R.}, \au{{Kawashima}, T.},
  \au{{Keating}, G.~K.}, \au{{Kettenis}, M.}, \au{{Kim}, J.-Y.}, \au{{Kim},
  J.}, \au{{Kim}, J.}, \au{{Kino}, M.}, \au{{Koay}, J.~Y.}, \au{{Koch}, P.~M.},
  \au{{Koyama}, S.}, \au{{Kramer}, M.}, \au{{Kramer}, C.}, \au{{Krichbaum},
  T.~P.}, \au{{Kuo}, C.-Y.}, \au{{Lauer}, T.~R.}, \au{{Lee}, S.-S.}, \au{{Li},
  Y.-R.}, \au{{Li}, Z.}, \au{{Lindqvist}, M.}, \au{{Liu}, K.}, \au{{Liuzzo},
  E.}, \au{{Lo}, W.-P.}, \au{{Lobanov}, A.~P.}, \au{{Loinard}, L.},
  \au{{Lonsdale}, C.}, \au{{Lu}, R.-S.}, \au{{MacDonald}, N.~R.}, \au{{Mao},
  J.}, \au{{Markoff}, S.}, \au{{Marrone}, D.~P.}, \au{{Marscher}, A.~P.},
  \au{{Mart{\'\i}-Vidal}, I.}, \au{{Matsushita}, S.}, \au{{Matthews}, L.~D.},
  \au{{Medeiros}, L.}, \au{{Menten}, K.~M.}, \au{{Mizuno}, Y.}, \au{{Mizuno},
  I.}, \au{{Moran}, J.~M.}, \au{{Moriyama}, K.}, \au{{Moscibrodzka}, M.},
  \au{{Mul{\ensuremath{\ddot{}}}ler}, C.}, \au{{Nagai}, H.}, \au{{Nagar},
  N.~M.}, \au{{Nakamura}, M.}, \au{{Narayan}, R.}, \au{{Narayanan}, G.},
  \au{{Natarajan}, I.}, \au{{Neri}, R.}, \au{{Ni}, C.}, \au{{Noutsos}, A.},
  \au{{Okino}, H.}, \au{{Olivares}, H.}, \au{{Oyama}, T.}, \au{{{\"O}zel}, F.},
  \au{{Palumbo}, D. C.~M.}, \au{{Patel}, N.}, \au{{Pen}, U.-L.}, \au{{Pesce},
  D.~W.}, \au{{Pi{\'e}tu}, V.}, \au{{Plambeck}, R.}, \au{{PopStefanija}, A.},
  \au{{Porth}, O.}, \au{{Prather}, B.}, \au{{Preciado-L{\'o}pez}, J.~A.},
  \au{{Psaltis}, D.}, \au{{Pu}, H.-Y.}, \au{{Ramakrishnan}, V.}, \au{{Rao},
  R.}, \au{{Rawlings}, M.~G.}, \au{{Raymond}, A.~W.}, \au{{Rezzolla}, L.},
  \au{{Ripperda}, B.}, \au{{Roelofs}, F.}, \au{{Rogers}, A.}, \au{{Ros}, E.},
  \au{{Rose}, M.}, \au{{Roshanineshat}, A.}, \au{{Rottmann}, H.}, \au{{Roy},
  A.~L.}, \au{{Ruszczyk}, C.}, \au{{Ryan}, B.~R.}, \au{{Rygl}, K. L.~J.},
  \au{{S{\'a}nchez}, S.}, \au{{S{\'a}nchez-Arguelles}, D.}, \au{{Sasada}, M.},
  \au{{Savolainen}, T.}, \au{{Schloerb}, F.~P.}, \au{{Schuster}, K.-F.},
  \au{{Shao}, L.}, \au{{Shen}, Z.}, \au{{Small}, D.}, \au{{Sohn}, B.~W.},
  \au{{SooHoo}, J.}, \au{{Tazaki}, F.}, \au{{Tiede}, P.}, \au{{Tilanus}, R.
  P.~J.}, \au{{Titus}, M.}, \au{{Toma}, K.}, \au{{Torne}, P.}, \au{{Trent},
  T.}, \au{{Trippe}, S.}, \au{{Tsuda}, S.}, \au{{van Bemmel}, I.}, \au{{van
  Langevelde}, H.~J.}, \au{{van Rossum}, D.~R.}, \au{{Wagner}, J.},
  \au{{Wardle}, J.}, \au{{Weintroub}, J.}, \au{{Wex}, N.}, \au{{Wharton}, R.},
  \au{{Wielgus}, M.}, \au{{Wong}, G.~N.}, \au{{Wu}, Q.}, \au{{Young}, A.},
  \au{{Young}, K.}, \au{{Younsi}, Z.}, \au{{Yuan}, F.}, \au{{Yuan}, Y.-F.},
  \au{{Zensus}, J.~A.}, \au{{Zhao}, G.}, \au{{Zhao}, S.-S.}, \au{{Zhu}, Z.},
  \au{{Anczarski}, J.}, \au{{Baganoff}, F.~K.}, \au{{Eckart}, A.}, \au{{Farah},
  J.~R.}, \au{{Haggard}, D.}, \au{{Meyer-Zhao}, Z.}, \au{{Michalik}, D.},
  \au{{Nadolski}, A.}, \au{{Neilsen}, J.}, \au{{Nishioka}, H.}, \au{{Nowak},
  M.~A.}, \au{{Pradel}, N.}, \au{{Primiani}, R.~A.}, \au{{Souccar}, K.},
  \au{{Vertatschitsch}, L.}, \au{{Yamaguchi}, P.} \& \au{{Zhang}, S.}}
  \yr{2019}  \at{{First M87 Event Horizon Telescope Results. V. Physical Origin
  of the Asymmetric Ring}}.  \jt{\apjl}  \bvol{875}~(1),  \pg{L5},
  \arxiv{arXiv: 1906.11242}.

\bibitem[{Galishnikova} {\em et~al.\/}(2023){Galishnikova}, {Philippov},
  {Quataert}, {Bacchini}, {Parfrey} \& {Ripperda}]{Galishnikova2023}
{\sc \au{{Galishnikova}, A.}, \au{{Philippov}, A.}, \au{{Quataert}, E.},
  \au{{Bacchini}, F.}, \au{{Parfrey}, K.} \& \au{{Ripperda}, B.}} \yr{2023}
  \at{{Collisionless Accretion onto Black Holes: Dynamics and Flares}}.
  \jt{\prl}  \bvol{130}~(11),  \pg{115201},  \arxiv{arXiv: 2212.02583}.

\bibitem[Gillies {\em et~al.\/}(2024)Gillies, van~der Wel, Van~den Bossche,
  Taves, Arnott, Ward {\em et~al.\/}]{shapely}
{\sc \au{Gillies, S.}, \au{van~der Wel, C.}, \au{Van~den Bossche, J.},
  \au{Taves, M.~W.}, \au{Arnott, J.}, \au{Ward, B.~C.} \& \au{others}}
  \yr{2024} Shapely.

\bibitem[{Goldreich} \& {Sridhar}(1995)]{GS95}
{\sc \au{{Goldreich}, P.} \& \au{{Sridhar}, S.}} \yr{1995}  \at{{Toward a
  Theory of Interstellar Turbulence. II. Strong Alfvenic Turbulence}}.
  \jt{\apj}  \bvol{438},  \pg{763}.

\bibitem[{Howes}(2010)]{Howes2010}
{\sc \au{{Howes}, G.~G.}} \yr{2010}  \at{{A prescription for the turbulent
  heating of astrophysical plasmas}}.  \jt{\mnras}  \bvol{409}~(1),
  \pg{L104--L108},  \arxiv{arXiv: 1009.4212}.

\bibitem[Howes(2016)]{howes_2016}
{\sc \au{Howes, G.~G.}} \yr{2016}  \at{The dynamical generation of current
  sheets in astrophysical plasma turbulence}.  \jt{The Astrophysical Journal
  Letters}  \bvol{827}~(2),  \pg{L28}.

\bibitem[{Kawazura} {\em et~al.\/}(2019){Kawazura}, {Barnes} \&
  {Schekochihin}]{Kawazura2019}
{\sc \au{{Kawazura}, Y.}, \au{{Barnes}, M.} \& \au{{Schekochihin}, A.~A.}}
  \yr{2019}  \at{{Thermal disequilibration of ions and electrons by
  collisionless plasma turbulence}}.  \jt{Proceedings of the National Academy
  of Science}  \bvol{116}~(3),  \pg{771--776},  \arxiv{arXiv: 1807.07702}.

\bibitem[{Kempski} {\em et~al.\/}(2023){Kempski}, {Fielding}, {Quataert},
  {Galishnikova}, {Kunz}, {Philippov} \& {Ripperda}]{Kempski2023}
{\sc \au{{Kempski}, P.}, \au{{Fielding}, D.~B.}, \au{{Quataert}, E.},
  \au{{Galishnikova}, A.~K.}, \au{{Kunz}, M.~W.}, \au{{Philippov}, A.~A.} \&
  \au{{Ripperda}, B.}} \yr{2023}  \at{{Cosmic ray transport in large-amplitude
  turbulence with small-scale field reversals}}.  \jt{\mnras}  \bvol{525}~(4),
  \pg{4985--4998},  \arxiv{arXiv: 2304.12335}.

\bibitem[{Kennel} \& {Coroniti}(1984{\natexlab{{\em a\/}}})]{Kennel84a}
{\sc \au{{Kennel}, C.~F.} \& \au{{Coroniti}, F.~V.}} \yr{1984{\natexlab{{\em
  a\/}}}}  \at{{Confinement of the Crab pulsar's wind by its supernova
  remnant.}}  \jt{\apj}  \bvol{283},  \pg{694--709}.

\bibitem[{Kennel} \& {Coroniti}(1984{\natexlab{{\em b\/}}})]{Kennel84b}
{\sc \au{{Kennel}, C.~F.} \& \au{{Coroniti}, F.~V.}} \yr{1984{\natexlab{{\em
  b\/}}}}  \at{{Magnetohydrodynamic model of Crab nebula radiation.}}
  \jt{\apj}  \bvol{283},  \pg{710--730}.

\bibitem[Kohonen(2001)]{Kohonen2001}
{\sc \au{Kohonen, T.}} \yr{2001} {\em Self-Organizing Maps\/}.  \publ{Springer
  Berlin Heidelberg}.

\bibitem[Kohonen(2013)]{Kohonen2013}
{\sc \au{Kohonen, T.}} \yr{2013}  \at{Essentials of the self-organizing map}.
  \jt{Neural Networks}  \bvol{37},  \pg{52–65}.

\bibitem[{Lemoine}(2023)]{Lemoine2023}
{\sc \au{{Lemoine}, M.}} \yr{2023}  \at{{Particle transport through localized
  interactions with sharp magnetic field bends in MHD turbulence}}.
  \jt{Journal of Plasma Physics}  \bvol{89}~(5),  \pg{175890501},
  \arxiv{arXiv: 2304.03023}.

\bibitem[Loureiro \& Boldyrev(2017)]{loureiro_boldyrev_2017}
{\sc \au{Loureiro, N.~F.} \& \au{Boldyrev, S.}} \yr{2017}  \at{Collisionless
  reconnection in magnetohydrodynamic and kinetic turbulence}.  \jt{The
  Astrophysical Journal}  \bvol{850}~(2),  \pg{182}.

\bibitem[{Luo} {\em et~al.\/}(2020){Luo}, {Lyutikov}, {Temim} \&
  {Comisso}]{Luo2020}
{\sc \au{{Luo}, Y.}, \au{{Lyutikov}, M.}, \au{{Temim}, T.} \& \au{{Comisso},
  L.}} \yr{2020}  \at{{Turbulent Model of Crab Nebula Radiation}}.  \jt{\apj}
  \bvol{896}~(2),  \pg{147},  \arxiv{arXiv: 2005.06319}.

\bibitem[Makwana {\em et~al.\/}(2015)Makwana, Zhdankin, Li, Daughton \&
  Cattaneo]{makwana_etal_2015}
{\sc \au{Makwana, K.}, \au{Zhdankin, V.}, \au{Li, H.}, \au{Daughton, W.} \&
  \au{Cattaneo, F.}} \yr{2015}  \at{Energy dynamics and current sheet structure
  in fluid and kinetic simulations of decaying magnetohydrodynamic turbulence}.
   \jt{Physics of Plasmas}  \bvol{22}~(4).

\bibitem[Mallet {\em et~al.\/}(2017)Mallet, Schekochihin \&
  Chandran]{mallet_etal_2017}
{\sc \au{Mallet, A.}, \au{Schekochihin, A.~A.} \& \au{Chandran, B.~D.}}
  \yr{2017}  \at{Disruption of alfv{\'e}nic turbulence by magnetic reconnection
  in a collisionless plasma}.  \jt{Journal of Plasma Physics}  \bvol{83}~(6),
  \pg{905830609}.

\bibitem[Meringolo {\em et~al.\/}(2023)Meringolo, Cruz-Osorio, Rezzolla \&
  Servidio]{meringolo_etal_2023}
{\sc \au{Meringolo, C.}, \au{Cruz-Osorio, A.}, \au{Rezzolla, L.} \&
  \au{Servidio, S.}} \yr{2023}  \at{Microphysical plasma relations from
  special-relativistic turbulence}.  \jt{The Astrophysical Journal}
  \bvol{944}~(2),  \pg{122}.

\bibitem[{Mo{\'s}cibrodzka} {\em et~al.\/}(2016){Mo{\'s}cibrodzka}, {Falcke} \&
  {Shiokawa}]{Moscibrodkzka2016}
{\sc \au{{Mo{\'s}cibrodzka}, M.}, \au{{Falcke}, H.} \& \au{{Shiokawa}, H.}}
  \yr{2016}  \at{{General relativistic magnetohydrodynamical simulations of the
  jet in M 87}}.  \jt{\aap}  \bvol{586},  \pg{A38},  \arxiv{arXiv: 1510.07243}.

\bibitem[N{\"a}ttil{\"a}(2022)]{nattila2022c}
{\sc \au{N{\"a}ttil{\"a}, J.}} \yr{2022}  \at{Runko: {{Modern}} multiphysics
  toolbox for plasma simulations}.  \jt{Astronomy \& Astrophysics}  \bvol{664},
   \pg{A68}.

\bibitem[N{\"a}ttil{\"a}(2024)]{nattila2024}
{\sc \au{N{\"a}ttil{\"a}, J.}} \yr{2024}  \at{Radiative plasma simulations of
  black hole accretion flow coronae in the hard and soft states}.  \jt{Nature
  Communications}  \bvol{15}~(1),  \pg{7026}.

\bibitem[{N{\"a}ttil{\"a}} \& {Beloborodov}(2021)]{2021ApJ...921...87N}
{\sc \au{{N{\"a}ttil{\"a}}, J.} \& \au{{Beloborodov}, A.~M.}} \yr{2021}
  \at{{Radiative Turbulent Flares in Magnetically Dominated Plasmas}}.
  \jt{\apj}  \bvol{921}~(1),  \pg{87},  \arxiv{arXiv: 2012.03043}.

\bibitem[N{\"a}ttil{\"a} \& Beloborodov(2022)]{nattila2022b}
{\sc \au{N{\"a}ttil{\"a}, J.} \& \au{Beloborodov, A.~M.}} \yr{2022}
  \at{Heating of {{Magnetically Dominated Plasma}} by {{Alfv\'en-Wave
  Turbulence}}}.  \jt{Physical Review Letters}  \bvol{128}~(7),  \pg{075101}.

\bibitem[Nurisso {\em et~al.\/}(2023)Nurisso, Celotti, Mignone \&
  Bodo]{nurisso_etal_2023}
{\sc \au{Nurisso, M.}, \au{Celotti, A.}, \au{Mignone, A.} \& \au{Bodo, G.}}
  \yr{2023}  \at{Particle acceleration with magnetic reconnection in
  large-scale rmhd simulations--i. current sheet identification and
  characterization}.  \jt{Monthly Notices of the Royal Astronomical Society}
  \bvol{522}~(4),  \pg{5517--5528}.

\bibitem[Parashar \& Matthaeus(2016)]{parashar_matthaeus_2016}
{\sc \au{Parashar, T.~N.} \& \au{Matthaeus, W.~H.}} \yr{2016}  \at{Propinquity
  of current and vortex structures: effects on collisionless plasma heating}.
  \jt{The Astrophysical Journal}  \bvol{832}~(1),  \pg{57}.

\bibitem[{Rees} \& {Gunn}(1974)]{Rees&Gunn74}
{\sc \au{{Rees}, M.~J.} \& \au{{Gunn}, J.~E.}} \yr{1974}  \at{{The origin of
  the magnetic field and relativistic particles in the Crab Nebula}}.
  \jt{\mnras}  \bvol{167},  \pg{1--12}.

\bibitem[{Ripperda} {\em et~al.\/}(2021){Ripperda}, {Mahlmann}, {Chernoglazov},
  {TenBarge}, {Most}, {Juno}, {Yuan}, {Philippov} \&
  {Bhattacharjee}]{tenbarge2021b}
{\sc \au{{Ripperda}, B.}, \au{{Mahlmann}, J.~F.}, \au{{Chernoglazov}, A.},
  \au{{TenBarge}, J.~M.}, \au{{Most}, E.~R.}, \au{{Juno}, J.}, \au{{Yuan}, Y.},
  \au{{Philippov}, A.~A.} \& \au{{Bhattacharjee}, A.}} \yr{2021}  \at{{Weak
  Alfv{\'e}nic turbulence in relativistic plasmas. Part 2. current sheets and
  dissipation}}.  \jt{Journal of Plasma Physics}  \bvol{87}~(5),
  \pg{905870512},  \arxiv{arXiv: 2105.01145}.

\bibitem[Ross \& Latter(2018)]{ross_latter_2018}
{\sc \au{Ross, J.} \& \au{Latter, H.~N.}} \yr{2018}  \at{Dissipative structures
  in magnetorotational turbulence}.  \jt{Monthly Notices of the Royal
  Astronomical Society}  \bvol{477}~(3),  \pg{3329--3342}.

\bibitem[{Ruszkowski} \& {Pfrommer}(2023)]{Ruszkowski2023}
{\sc \au{{Ruszkowski}, M.} \& \au{{Pfrommer}, C.}} \yr{2023}  \at{{Cosmic ray
  feedback in galaxies and galaxy clusters}}.  \jt{\aapr}  \bvol{31}~(1),
  \pg{4},  \arxiv{arXiv: 2306.03141}.

\bibitem[{Schekochihin}(2022)]{schekochihin_2022}
{\sc \au{{Schekochihin}, A.~A.}} \yr{2022}  \at{{MHD turbulence: a biased
  review}}.  \jt{Journal of Plasma Physics}  \bvol{88}~(5),  \pg{155880501}.

\bibitem[Servidio {\em et~al.\/}(2010)Servidio, Matthaeus, Shay, Dmitruk,
  Cassak \& Wan]{servidio_etal_2010}
{\sc \au{Servidio, S.}, \au{Matthaeus, W.}, \au{Shay, M.}, \au{Dmitruk, P.},
  \au{Cassak, P.} \& \au{Wan, M.}} \yr{2010}  \at{Statistics of magnetic
  reconnection in two-dimensional magnetohydrodynamic turbulence}.  \jt{Physics
  of Plasmas}  \bvol{17}~(3).

\bibitem[Singh {\em et~al.\/}(2024)Singh, French, Guo \& Li]{singh_etal_2024}
{\sc \au{Singh, D.}, \au{French, O.}, \au{Guo, F.} \& \au{Li, X.}} \yr{2024}
  \at{Low-energy injection and nonthermal particle acceleration in relativistic
  magnetic turbulence}.  \jt{arXiv preprint arXiv:2404.19181} .

\bibitem[Sisti {\em et~al.\/}(2021)Sisti, Fadanelli, Cerri, Faganello, Califano
  \& Agullo]{sisti_etal_2021}
{\sc \au{Sisti, M.}, \au{Fadanelli, S.}, \au{Cerri, S.~S.}, \au{Faganello, M.},
  \au{Califano, F.} \& \au{Agullo, O.}} \yr{2021}  \at{Characterizing current
  structures in 3d hybrid-kinetic simulations of plasma turbulence}.
  \jt{Astronomy \& Astrophysics}  \bvol{655},  \pg{A107}.

\bibitem[TenBarge \& Howes(2013)]{tenbarge_etal_2013}
{\sc \au{TenBarge, J.~M.} \& \au{Howes, G.}} \yr{2013}  \at{Current sheets and
  collisionless damping in kinetic plasma turbulence}.  \jt{The Astrophysical
  Journal Letters}  \bvol{771}~(2),  \pg{L27}.

\bibitem[Umbach \& Jones(2003)]{Umbach2003}
{\sc \au{Umbach, D.} \& \au{Jones, K.}} \yr{2003}  \at{A few methods for
  fitting circles to data}.  \jt{IEEE Transactions on Instrumentation and
  Measurement}  \bvol{52}~(6),  \pg{1881--1885}.

\bibitem[Uritsky {\em et~al.\/}(2010)Uritsky, Pouquet, Rosenberg, Mininni \&
  Donovan]{uritsky_etal_2010}
{\sc \au{Uritsky, V.~M.}, \au{Pouquet, A.}, \au{Rosenberg, D.}, \au{Mininni,
  P.~D.} \& \au{Donovan, E.}} \yr{2010}  \at{Structures in magnetohydrodynamic
  turbulence: Detection and scaling}.  \jt{Physical Review E}  \bvol{82}~(5),
  \pg{056326}.

\bibitem[Vega {\em et~al.\/}(2022)Vega, Boldyrev, Roytershteyn \&
  Medvedev]{vega_etal_2022}
{\sc \au{Vega, C.}, \au{Boldyrev, S.}, \au{Roytershteyn, V.} \& \au{Medvedev,
  M.}} \yr{2022}  \at{Turbulence and particle acceleration in a relativistic
  plasma}.  \jt{The Astrophysical Journal Letters}  \bvol{924}~(1),  \pg{L19}.

\bibitem[Virtanen {\em et~al.\/}(2020)Virtanen, Gommers, Oliphant, Haberland,
  Reddy, Cournapeau, Burovski, Peterson, Weckesser, Bright, {van der Walt},
  Brett, Wilson, Millman, Mayorov, Nelson, Jones, Kern, Larson, Carey, Polat,
  Feng, Moore, {VanderPlas}, Laxalde, Perktold, Cimrman, Henriksen, Quintero,
  Harris, Archibald, Ribeiro, Pedregosa, {van Mulbregt} \& {SciPy 1.0
  Contributors}]{2020SciPy-NMeth}
{\sc \au{Virtanen, P.}, \au{Gommers, R.}, \au{Oliphant, T.~E.}, \au{Haberland,
  M.}, \au{Reddy, T.}, \au{Cournapeau, D.}, \au{Burovski, E.}, \au{Peterson,
  P.}, \au{Weckesser, W.}, \au{Bright, J.}, \au{{van der Walt}, S.~J.},
  \au{Brett, M.}, \au{Wilson, J.}, \au{Millman, K.~J.}, \au{Mayorov, N.},
  \au{Nelson, A. R.~J.}, \au{Jones, E.}, \au{Kern, R.}, \au{Larson, E.},
  \au{Carey, C.~J.}, \au{Polat, {\.I}.}, \au{Feng, Y.}, \au{Moore, E.~W.},
  \au{{VanderPlas}, J.}, \au{Laxalde, D.}, \au{Perktold, J.}, \au{Cimrman, R.},
  \au{Henriksen, I.}, \au{Quintero, E.~A.}, \au{Harris, C.~R.}, \au{Archibald,
  A.~M.}, \au{Ribeiro, A.~H.}, \au{Pedregosa, F.}, \au{{van Mulbregt}, P.} \&
  \au{{SciPy 1.0 Contributors}}} \yr{2020}  \at{{{SciPy} 1.0: Fundamental
  Algorithms for Scientific Computing in Python}}.  \jt{Nature Methods}
  \bvol{17},  \pg{261--272}.

\bibitem[Vlahos \& Isliker(2023)]{vlahos_etal_2023}
{\sc \au{Vlahos, L.} \& \au{Isliker, H.}} \yr{2023}  \at{Formation and
  evolution of coherent structures in 3d strongly turbulent magnetized
  plasmas}.  \jt{Physics of Plasmas}  \bvol{30}~(4).

\bibitem[Wan {\em et~al.\/}(2012)Wan, Matthaeus, Karimabadi, Roytershteyn,
  Shay, Wu, Daughton, Loring \& Chapman]{wan_etal_2012}
{\sc \au{Wan, M.}, \au{Matthaeus, W.}, \au{Karimabadi, H.}, \au{Roytershteyn,
  V.}, \au{Shay, M.}, \au{Wu, P.}, \au{Daughton, W.}, \au{Loring, B.} \&
  \au{Chapman, S.~C.}} \yr{2012}  \at{Intermittent dissipation at kinetic
  scales in collisionless plasma turbulence}.  \jt{Physical review letters}
  \bvol{109}~(19),  \pg{195001}.

\bibitem[{Werner} {\em et~al.\/}(2018){Werner}, {Uzdensky}, {Begelman},
  {Cerutti} \& {Nalewajko}]{Werner2018}
{\sc \au{{Werner}, G.~R.}, \au{{Uzdensky}, D.~A.}, \au{{Begelman}, M.~C.},
  \au{{Cerutti}, B.} \& \au{{Nalewajko}, K.}} \yr{2018}  \at{{Non-thermal
  particle acceleration in collisionless relativistic electron-proton
  reconnection}}.  \jt{\mnras}  \bvol{473}~(4),  \pg{4840--4861},
  \arxiv{arXiv: 1612.04493}.

\bibitem[Zhdankin {\em et~al.\/}(2014)Zhdankin, Boldyrev, Perez \&
  Tobias]{zhdankin_etal_2014}
{\sc \au{Zhdankin, V.}, \au{Boldyrev, S.}, \au{Perez, J.~C.} \& \au{Tobias,
  S.~M.}} \yr{2014}  \at{Energy dissipation in magnetohydrodynamic turbulence:
  coherent structures or “nanoflares”?}  \jt{The Astrophysical Journal}
  \bvol{795}~(2),  \pg{127}.

\bibitem[{Zhdankin} {\em et~al.\/}(2016){Zhdankin}, {Boldyrev} \&
  {Uzdensky}]{zhdankin_etal_2016}
{\sc \au{{Zhdankin}, V.}, \au{{Boldyrev}, S.} \& \au{{Uzdensky}, D.~A.}}
  \yr{2016}  \at{{Scalings of intermittent structures in magnetohydrodynamic
  turbulence}}.  \jt{Physics of Plasmas}  \bvol{23}~(5),  \pg{055705},
  \arxiv{arXiv: 1602.05289}.

\bibitem[{Zhdankin} {\em et~al.\/}(2013){Zhdankin}, {Uzdensky}, {Perez} \&
  {Boldyrev}]{2013ApJ...771..124Z}
{\sc \au{{Zhdankin}, V.}, \au{{Uzdensky}, D.~A.}, \au{{Perez}, J.~C.} \&
  \au{{Boldyrev}, S.}} \yr{2013}  \at{{Statistical Analysis of Current Sheets
  in Three-dimensional Magnetohydrodynamic Turbulence}}.  \jt{\apj}
  \bvol{771}~(2),  \pg{124},  \arxiv{arXiv: 1302.1460}.

\bibitem[{Zhdankin} {\em et~al.\/}(2017{\natexlab{{\em a\/}}}){Zhdankin},
  {Walker}, {Boldyrev} \& {Lesur}]{2017MNRAS.467.3620Z}
{\sc \au{{Zhdankin}, V.}, \au{{Walker}, J.}, \au{{Boldyrev}, S.} \&
  \au{{Lesur}, G.}} \yr{2017{\natexlab{{\em a\/}}}}  \at{{Universal small-scale
  structure in turbulence driven by magnetorotational instability}}.
  \jt{\mnras}  \bvol{467}~(3),  \pg{3620--3627},  \arxiv{arXiv: 1702.02857}.

\bibitem[{Zhdankin} {\em et~al.\/}(2017{\natexlab{{\em b\/}}}){Zhdankin},
  {Werner}, {Uzdensky} \& {Begelman}]{Zhdankin2017_prl}
{\sc \au{{Zhdankin}, V.}, \au{{Werner}, G.~R.}, \au{{Uzdensky}, D.~A.} \&
  \au{{Begelman}, M.~C.}} \yr{2017{\natexlab{{\em b\/}}}}  \at{{Kinetic
  Turbulence in Relativistic Plasma: From Thermal Bath to Nonthermal
  Continuum}}.  \jt{\prl}  \bvol{118}~(5),  \pg{055103},  \arxiv{arXiv:
  1609.04851}.

\end{thebibliography}

\makeatletter
\def\fps@table{h}
\def\fps@figure{h}
\makeatother




\end{document}